\documentclass[conference]{IEEEtran}
\ifCLASSINFOpdf
\else
\fi
\usepackage{amsmath}
\usepackage{xspace}
\usepackage{booktabs} 
\usepackage{makecell}
\usepackage{graphicx}
\usepackage{multirow}
\usepackage[dvipsnames]{xcolor}
\usepackage[utf8]{inputenc}
\usepackage{tcolorbox}
\usepackage{tikz}
\usepackage{enumitem}
\usepackage{hyperref}
\usepackage{subcaption}

\newcommand{\SysName}{{\scshape AEAS}\xspace}
\newcommand{\norm}[1]{\left\lVert #1 \right\rVert}

\newcommand{\XS}[1]{{#1}}

\begin{document}
%
\title{\SysName: Actionable Exploit Assessment System}



	

%
\author{\IEEEauthorblockN{Xiangmin Shen\IEEEauthorrefmark{1},
Wenyuan Cheng\IEEEauthorrefmark{2},
Yan Chen\IEEEauthorrefmark{3}, 
Zhenyuan Li\IEEEauthorrefmark{2}, \\
Yuqiao Gu\IEEEauthorrefmark{2}, 
Lingzhi Wang\IEEEauthorrefmark{3},
Wencheng Zhao\IEEEauthorrefmark{4},
Dawei Sun\IEEEauthorrefmark{4} and
Jiashui Wang\IEEEauthorrefmark{2}}
\IEEEauthorblockA{\IEEEauthorrefmark{1}Hofstra University, \IEEEauthorrefmark{2}Zhejiang University, \IEEEauthorrefmark{3}Northwestern University, \IEEEauthorrefmark{4}Ant Group}
}




\maketitle

\begin{abstract}
Security practitioners face growing challenges in exploit assessment, as public vulnerability repositories are increasingly populated with inconsistent and low-quality exploit artifacts. Existing scoring systems such as CVSS and EPSS offer limited support for this task. They either rely on theoretical metrics or produce opaque probability estimates without assessing whether usable exploit code exists. In practice, security teams often resort to manual triage of exploit repositories, which is time-consuming, error-prone, and difficult to scale.

We present \SysName, an automated system designed to assess and prioritize actionable exploits through static analysis. \SysName analyzes both exploit code and associated documentation to extract a structured set of features reflecting exploit availability, functionality, and setup complexity. It then computes an actionability score for each exploit and produces ranked exploit recommendations. 

We evaluate \SysName on a dataset of over 5,000 vulnerabilities derived from 600+ real-world applications frequently encountered by red teams. Manual validation and expert review on representative subsets show that \SysName achieves a 100\% top-3 success rate in recommending functional exploits and shows strong alignment with expert-validated rankings. These results demonstrate the effectiveness of \SysName in supporting exploit-driven vulnerability prioritization.
\end{abstract}


%
\IEEEpeerreviewmaketitle

\section{Introduction}

The ever-growing volume of cataloged vulnerabilities poses a significant challenge for security practitioners: \textit{how to effectively prioritize vulnerabilities for exploitation or remediation?}
Within the vast pool of known vulnerabilities, only a small subset is practically exploitable. Identifying these high-impact cases remains difficult due to the lack of actionable signals. Although severity scoring systems exist, they primarily focus on theoretical risk rather than practical exploitability. This gap has led to inefficiencies in both offensive operations and vulnerability triage workflows.

This challenge becomes more severe at scale. Automated scanners and monitoring systems routinely report large numbers of vulnerabilities, many of which have public exploit artifacts available. However, the substantial quantity of available exploits is accompanied by wide variation in quality. This inconsistency complicates efforts to identify which vulnerabilities are realistically exploitable in a given context. In our empirical study detailed in Section~\ref{subsec:eval_rq1}, we manually examined over 600 exploit candidates associated with 47 vulnerabilities and found that only 51.4\% were capable of achieving their intended impact. The remaining exploits were either non-functional, incomplete, or usable only as proof-of-concept (PoC) demonstrations. This disconnect between the presence of exploit artifacts and their actual utility underscores the limitations of current vulnerability assessment practices.



Current scoring systems provide limited support for this need. The Common Vulnerability Scoring System (CVSS)~\cite{cvss} and the Exploit Prediction Scoring System (EPSS)~\cite{epss} are among the most widely used. CVSS relies on manually assigned metrics that are often incomplete, or inconsistent across different reporters~\cite{wunder2024shedding, zhang2023flaw}. These scores reflect static characteristics of vulnerabilities but fail to account for real-world exploitation evidence. As a result, CVSS scores are known to correlate poorly with actual exploitation in the wild~\cite{allodi2014comparing, jacobs2023enhancing}. EPSS improves on this by incorporating external data sources and predictive modeling, but it lacks transparency, offering exploitation probability without evidence or reasoning led to the prediction. Both systems do not evaluate whether a concrete, runnable exploit exists or how usable it is in practice. As a result, practitioners often face a disconnect between high-severity scores and the operational reality of exploit availability. For instance, many vulnerabilities with high CVSS scores lack usable exploit code, while others with low scores may have functional exploits readily available in public repositories.


In parallel, penetration testing tools and vulnerability scanners have improved in indexing exploit-related information, but they rarely offer systematic prioritization grounded in exploit artifacts. Platforms~\cite{tenable,cvemap} such as Metasploit~\cite{metasploit} and Nuclei~\cite{nuclei} provide access to PoCs and exploit scripts. However, these tools often emphasize breadth over quality. Many listed PoCs are incomplete, difficult to execute, or require substantial modification before use. Moreover, few tools distinguish between verified exploits and scripts that merely reproduce conditions for triggering a vulnerability. As a result, practitioners must manually examine each artifact and determine whether it applies to their target environment. This process is labor-intensive and does not scale well.

Together, these limitations reveal two key gaps in the current vulnerability assessment landscape:

\begin{enumerate}[label=G\arabic*., left=0pt, itemsep=0em]

\item \underline{Actionability.} Practitioners lack systems that help identify and prioritize exploits that are truly ready for use. We define actionable exploits as those that are (i) available, (ii) functional, and (iii) require minimal setup. Availability ensures that an exploit can be obtained and inspected. Functionality determines whether the exploit can achieve meaningful impact, such as remote code execution or privilege escalation. Minimal setup reduces friction by minimizing environment-specific dependencies or the need for manual configuration. Each of these dimensions is necessary for an exploit to be useful in automated red teaming, exploit development, or remediation prioritization. Existing tools and scoring systems do not evaluate these properties, resulting in mismatches between perceived and actual risk.

\item \underline{Automation.} Most workflows still rely heavily on manual effort to assess exploit actionability. Analysts must examine exploit code, cross-reference documentation, and test scripts under different configurations. Although some tools automate the retrieval of exploit artifacts, few support automated reasoning across heterogeneous, unstructured inputs such as blog posts, README files, code snippets, or configuration instructions.

\end{enumerate}


Recent advances in large language models (LLMs) offer new opportunities for addressing these challenges. LLMs have demonstrated strong performance in tasks involving text summarization, contextual interpretation, and code analysis. These capabilities make them well-suited for analyzing unstructured security artifacts from sources such as public exploit repositories, and technical documentation. Early research has explored using LLMs to process unstructured security-related data like blogs and emails~\cite{ashiwal2024llm}. However, these efforts have generally focused on textual classification or summarization tasks and do not extend to exploit-level analysis. More importantly, they do not attempt to assess whether an exploit is available, functional, or easy-to-use.

In this paper, we present \SysName, an automated and actionable exploit assessment system designed to address a critical and underexplored problem: identifying and prioritizing vulnerabilities based on the actionability of public exploit artifacts. Rather than replacing existing systems such as CVSS or EPSS, \SysName complements them by answering several operational questions: whether a runnable exploit exists, whether it can be expected to work reliably in practice, and how difficult it is to deploy in real-world conditions.

A straightforward approach to this problem would be to execute each public exploit in a test environment and observe its behavior. However, this strategy is expensive, time-consuming, and difficult to scale. Many exploits require custom environments, manual configuration, or non-trivial setup. Large-scale execution also raises practical concerns, such as infrastructure cost, long test cycles, and difficulty maintaining a low profile in sensitive settings. As a result, we design \SysName to work through static analysis. Our goal is to approximate the benefits of dynamic analysis, understanding exploit feasibility and operational impact, without actually running the code.

This static analysis approach presents several technical challenges. Public exploit artifacts are noisy and highly inconsistent. Exploits often lack structured metadata and may be distributed alongside unstructured notes, blog posts, or incomplete documentation. To address this, \SysName includes a preprocessing pipeline that extracts contextual information, normalizes formats, and identifies relevant components for analysis. More importantly, exploit actionability cannot be captured through surface-level signals. It involves reasoning about various distinct features that contribute to exploit actionability, such as execution complexity and impact. We define a structured schema of such features and extract them using a hybrid approach that combines LLM-based reasoning and summarization with rule-based validation and aggregation. 

Extracting these features accurately is not a trivial application of existing LLM capabilities. Off-the-shelf prompting often yields noisy or inconsistent outputs, particularly for technical security tasks. To improve robustness, we introduce task-specific prompt templates and chain-of-thought reasoning to guide LLMs through structured interpretation. Our ablation study shows that these techniques improve both accuracy and consistency across diverse model backbones. The extracted features are used to compute an exploit-level actionability score, which is then aggregated into vulnerability-level rankings to reflect overall severity.

We construct a diverse evaluation pool of over 5,000 vulnerabilities published recently, covering a wide range of severities and exploit types. These vulnerabilities are linked to more than 600 applications frequently encountered by industry red teams, based on direct input from security professionals. We use \SysName to analyze this entire pool, demonstrating its scalability and broad applicability. We also conduct further manual verification and expert validation on representative subsets. The results show that \SysName achieves a 100\% success rate in its top-3 exploit recommendations and produces vulnerability-level rankings that closely match expert judgments, validating both the effectiveness and practical utility of the system.


\noindent\textbf{Contributions. } This paper makes the following contributions:

\begin{list}{\labelitemi}{\leftmargin=.5em}
 \setlength{\itemsep}{0em}
 \setlength{\parskip}{0pt}
 \setlength{\parsep}{0pt}
    \item We present \SysName, an automated framework for actionable exploit assessment. \SysName identifies high-quality exploits based on their functional and operational characteristics.~\footnote{\SysName has been implemented in the planning agent of \href{https://dl.acm.org/doi/10.1145/3708821.3733882}{PentestAgent}, our LLM-driven automated penetration testing framework. Check out our repository: https://github.com/nbshenxm/pentest-agent}
    \item \SysName introduces a task-specific analysis pipeline that combines large language models with rule-based heuristics to extract and reason over relevant features from unstructured exploit artifacts, enabling accurate analysis without requiring dynamic execution. To our knowledge, this is the first framework tailored for structured exploit assessment at scale.
    \item We evaluate \SysName on over 5,000 real-world vulnerabilities derived from commonly encountered applications in professional red teaming. The system achieves a 100\% top-3 success rate in exploit recommendation and strong agreement with expert rankings, demonstrating its effectiveness and practical value.
\end{list}




\section{Background and Related Work}\label{sec:background}
\subsection{Vulnerability Assessment}
Vulnerability assessment is a critical process for identifying and prioritizing vulnerabilities within a system, especially when dealing with a large volume of assets. With numerous potential vulnerabilities across a network or environment, evaluating which ones pose the greatest risk is a complex task. Without a structured analysis, it is inefficient and impractical to address every vulnerability individually. Instead, vulnerability assessment provides a structured analysis from multiple perspectives to help testers prioritize vulnerabilities to exploit effectively.
Over the years, several security metrics have been developed to assess the vulnerabilities~\cite{milousi2024evaluating}, with the CVSS\cite{cvss} and the EPSS\cite{epss} being the most widely adopted for system security assessments.

CVSS assigns a severity score ranging from 0.0 to 10.0 based on factors such as exploitability and impact. While widely used, CVSS suffers from significant issues of accuracy and consistency~\cite{zhang2023flaw}. Wunder et al.~\cite{wunder2024shedding} highlighted key shortcomings in a user study, pointing to ambiguous metric definitions and arbitrary manual scoring, often conducted without proper adherence to CVSS documentation. These inconsistencies undermine its reliability as a prioritization tool.
A significant enhancement in CVSS v4.0 is the exploit maturity metric, which estimates the likelihood of a vulnerability being exploited based on factors such as exploit techniques, the availability of exploit code, and active ``in-the-wild" exploitation~\cite{cvss_exploitability}. Despite its potential value, this metric is rarely included in published CVSS scores, leaving a critical gap in practical vulnerability prioritization.

In response to these gaps, several external sources have been developed to provide insights into exploiting maturity. For example, EPSS~\cite{jacobs2023enhancing} employs a predictive model to estimate the likelihood of a vulnerability being exploited. While EPSS introduces a useful predictive capability, it lacks transparency in its outputs, making it difficult for practitioners to understand or validate the rationale behind the results.
Several academic works~\cite{jacobs2020improving, le2022survey, yin2023empowering, fang2020fastembed, harzevili2023survey, zhang2023flaw} aim to improve vulnerability assessment by leveraging machine learning techniques, such as NLP techniques~\cite{suciu2022expected} and neural networks~\cite{yin2023empowering}. However, like EPSS, these approaches still suffer from significant challenges related to transparency and explainability, limiting their practical utility in real-world scenarios.

Commercial vulnerability management tools also attempt to address these shortcomings by integrating exploit maturity assessments into their services. 
For instance, Tenable~\cite{tenable} offers a Vulnerability Priority Rating (VPR), providing a more nuanced and sophisticated metric for vulnerability prioritization. However, these tools have significant drawbacks: full access to their vulnerability databases and advanced assessment features often require costly subscriptions, creating barriers for smaller organizations or independent researchers. Moreover, the proprietary nature of these methods limits transparency, making it difficult to assess the reliability or fairness of their evaluations.
Besides, Community-driven initiatives, such as CVEmap~\cite{cvemap}, aggregate information from public sources to improve accessibility. However, these projects tend to focus primarily on data collection without deeper analysis, which limits their effectiveness in supporting vulnerability prioritization.

These challenges underscore the need for an automated and explainable vulnerability assessment system. Such a system could improve both accuracy and consistency, while offering clear, transparent reasoning behind its outputs, providing practitioners with actionable insights backed by transparent decision-making processes.

\subsection{Exploit Assessment and Recommendation}

Effective exploit assessment is crucial for identifying which available exploits are suitable for executing attacks. When vulnerabilities are discovered, a wide range of exploits or proofs of concept (PoCs) are often associated with them. However, not all of these exploits or PoCs are of equal quality or applicability. Sometimes, it claims to be an exploit but, in fact, a PoC and verse visa. Therefore, determining the usability of exploits is crucial. Effective exploit assessment should consider factors such as exploit reliability, execution feasibility, and contextual applicability to the target environment.

Metasploit~\cite{metasploit} is a widely used exploitation framework that provides a collection of ready-to-use exploits. Metasploit includes an exploit ranking system that helps testers quickly identify practical attack methods. However, Metasploit has several limitations. Its repository contains a relatively small number of exploits, meaning it may not support newly discovered vulnerabilities in a timely manner. Additionally, the exploit rankings provided by Metasploit~\cite{metasploit_exploit_ranking} lack transparency, offering no detailed explanation for why a particular exploit is ranked as more feasible or effective. Lastly, the exploit ranking only considers the reliability.

Community-driven resources, such as Nuclei~\cite{nuclei}, provide comprehensive repositories for vulnerability-related information. These platforms are known for their rapid updates and broad coverage of PoCs. However, they focus primarily on PoC availability. This lack of coverage on available exploits leaves penetration testers with the responsibility of manually searching for exploits, as well as evaluating and selecting the most suitable exploit, which can be time-consuming and error-prone. 

Several academic works~\cite{fedorchenko2023analytical, suciu2022expected} study the exploit code structure in attempt to gain insights. Suciu et al.~\cite{suciu2022expected} attempted to improve exploitability prediction by analyzing PoCs. While this approach is helpful for vulnerability prioritization, it does not provide a systematic method for assessing the usability of existing exploits, leaving a significant gap in the exploitation assessment process.

Given these limitations, there is a clear need for an automated, explainable exploit assessment and recommendation system. Such a system would analyze various factors influencing exploit usability, including exploit characteristics, reliability, and environmental constraints, enabling it to rank exploits effectively.





\subsection{LLM-based Data Analysis}

Building on the need for efficient vulnerability and exploit assessment, large language models (LLMs) present a promising solution for analyzing unstructured data. LLMs have demonstrated exceptional capabilities in fields like finance~\cite{li2023extracting} and healthcare~\cite{wiest2024llm, wang2025survey, lin2025healthgpt}, where their ability to process and summarize vast amounts of text and interpret domain-specific information has significantly improved decision-making processes. Similarly, vulnerability and exploit assessment can leverage LLMs to extract valuable insights from diverse unstructured data sources, including technical blogs, vulnerability disclosures, exploit code, and software documentation.

Initial research efforts~\cite{fayyazi2023uses} have explored the application of LLMs in this context. For instance, Ashiwal et al.~\cite{ashiwal2024llm} proposed a framework that utilizes LLMs to process unstructured data, such as blogs and emails, and generate structured outputs relevant to security analysis. While this work highlights the potential of LLMs in aiding penetration testing workflows, it is limited to text-based analysis and does not fully consider the variety of data formats encountered in penetration testing, such as code snippets, configuration files, or exploit repositories.
Several works have achieved some initial successes in detecting vulnerabilities~\cite{purba2023software, jensen2024software, zhou2024large} from the source code and generating patches~\cite{ahmed2023better}.
These successes make us think we can use LLM to fill the gap of exploit assessment.

Despite their potential, generic LLMs face significant challenges in effectively supporting vulnerability and exploit assessment tasks. First, LLMs do not inherently understand what specific features are relevant to penetration testers. Without clear guidance, LLMs may produce outputs that lack actionable insights or omit critical details. Second, even when given instructions to focus on specific features, LLMs often struggle to extract relevant information effectively and efficiently. This is due to their reliance on pre-trained general knowledge rather than utilizing the provided external context optimally. For example, an LLM may incorrectly attempt to extract features from documentation when the answer lies in associated code, or vice versa. Such inefficiencies not only hinder the assessment process but can also lead to hallucination issues, where the model generates incorrect or misleading information.

Many LLM techniques can help address these challenges.
Retrieval-Augmented Generation (RAG)~\cite{lewis2020retrieval} enhances LLMs by allowing them to utilize external data for generating responses. This technique involves three main stages: indexing, retrieval, and response synthesis. Initially, the dataset is indexed for efficient retrieval. Upon receiving a query, RAG retrieves relevant information from the indexed dataset and combines it with the original query before sending it to the LLM for response synthesis. 
The chain-of-thought (CoT)~\cite{wei2022chain} technique significantly improves the ability of large language models to perform complex reasoning. By guiding the LLM to follow a logical sequence of steps, this method enhances the model's problem-solving capabilities.
Role-playing~\cite{li2023camel} asks the LLM to impersonate an imaginary character, allowing the LLM to operate with clear objectives and boundaries, thereby enhancing their efficiency and effectiveness.
Structured output techniques can save time spent on iterative prompt testing and ad-hoc parsing, reducing overall LLM inference costs and latency, as well as developers' effort. Additionally, structured outputs ensure smooth integration with downstream processes and workflows~\cite{liu2024we}.

Using these techniques collectively, we propose the design of a customized data processing pipeline tailored to the specific needs of penetration testing. This pipeline leverages a combination of techniques to guide the LLM in extracting relevant features from various data sources, ensuring a more targeted and reliable analysis. By incorporating domain-specific instructions and seamlessly handling multiple data formats, our approach enhances the precision and effectiveness of LLM-based vulnerability and exploit assessments. This advancement bridges the gap between general-purpose LLM capabilities and the specialized requirements of exploit assessment, contributing to more accurate, transparent, and actionable insights.

\section{System Design}\label{sec:design}
\subsection{System Overview}
\begin{figure}[!htbp]
\vspace{-0.5em}
    \centering
    \includegraphics[width=1.0\linewidth]{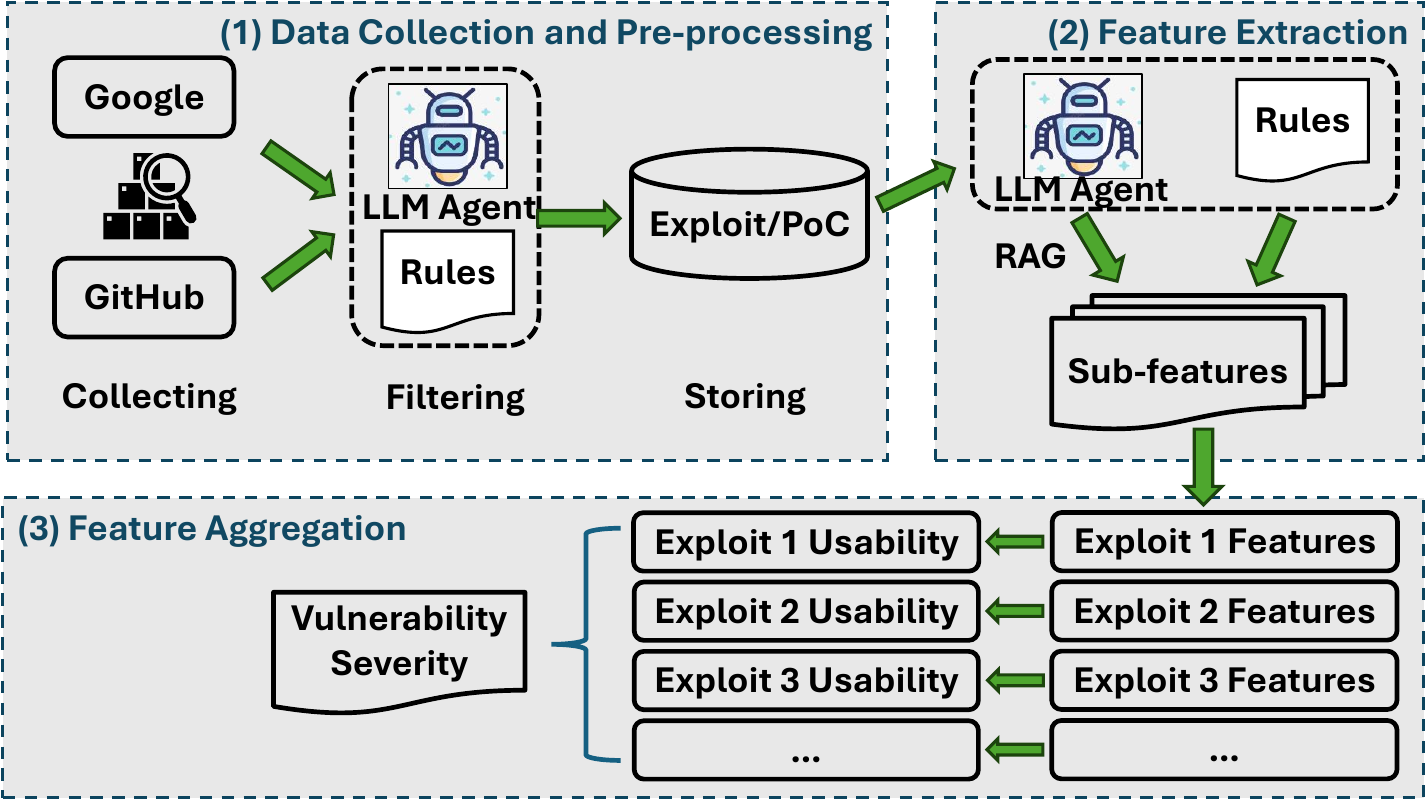}
    \caption{System Overview}
    \label{fig:system_overview}
\end{figure}
As shown in Fig.~\ref{fig:system_overview}, \SysName consists of three main stages: (1) Data Collection and Pre-processing, (2) Feature Extraction, and (3) Feature Aggregation. 
At the data collection stage, we collect data from designated data sources and apply both rule-based and LLM-driven filtering to facilitate more efficient processing in the subsequent stages. 
During the Feature Extraction stage, an LLM agent and rule-based heuristics are employed to extract sub-features, which are then combined to form comprehensive features for each exploit.
Finally, in the Feature Aggregation stage, these features are used to compute exploit exploitability scores, which are subsequently aggregated to determine overall vulnerability severity.
We will further elucidate each stage in Section~\ref{subsec:data_collection}, ~\ref{subsec:feature_extraction}, and ~\ref{subsec:feature_aggr}, respectively.

\subsection{Data Collection and Pre-processing}
\label{subsec:data_collection}

\XS{

We collect vulnerability-related data from Google and GitHub, two widely used and complementary sources in real-world security workflows. Google enables access to web-based documentation, such as blog posts, tutorials, technical Q\&A, and exploitation writeups that describe conditions for successful exploitation or mitigation. GitHub hosts a large volume of vulnerability-related repositories, including PoC scripts, exploit frameworks, and setup guides. Together, these sources provide both high-level contextual information and low-level technical artifacts relevant to vulnerability and exploit assessment.

However, data retrieved from these platforms is highly heterogeneous and noisy. Search results often include duplicated CVE summaries, advertisements, incomplete code templates, and repositories lacking functional content. These inconsistencies complicate downstream analysis and introduce irrelevant or misleading input.

To address these issues, we develop a hybrid pre-processing pipeline that combines rule-based heuristics with LLM agents. The rule-based component eliminates clearly uninformative or malformed data (e.g., binary files, enumerations of vulnerabilities, irrelevant metadata), while the LLM agent is used to analyze remaining content and filter out semantically trivial material, such as boilerplate summaries, generic explanations, and incomplete exploits.

This joint approach improves robustness against the diverse and unstructured nature of the input data. Rule-based filtering ensures efficiency and coverage for easily detectable artifacts, while the LLM enables semantic understanding across varied formats and documentation styles. Together, these techniques allow us to retain relevant content and discard noise, ensuring that subsequent analysis modules operate on meaningful and actionable inputs. Further details of the data pre-processing steps are provided in Appendix~\ref{appendix:github_filtering}.

}

\begin{table}[!htbp]
\centering

\caption{Key Features and Corresponding Sub-Features}

\label{tab:vul_features}
\begin{tabular}{c|c}

\toprule
Feature      & Sub-Feature    \\ \midrule
Attack Vector & IsRemote \\ \midrule
Attack Complexity & \makecell{Info. Dependency, Attack Condition, \\Probability, UI, Privilege Req., Evasion}  \\ \midrule
Impact & \makecell{Code Exec., Privilege Escalation, \\Info. Leak, Bypass, DoS}\\ \midrule
Exploit Maturity & \makecell{Relevance, Availability, \\Flexibility, Functionality} \\ \midrule
Popularity & \# of Exploits, \# of GitHub Stars \& Forks  \\ 

\bottomrule
\end{tabular}
\end{table}

\subsection{Feature Extraction}
\label{subsec:feature_extraction}

After completing data collection and preprocessing, we analyze each exploit along a structured set of features designed to capture its actionability. These features, summarized in Table~\ref{tab:vul_features}, span five dimensions: attack vector, attack complexity, impact, exploit maturity, and popularity. Each dimension is further broken down into sub-features to support fine-grained assessment.

Our feature design is grounded in two foundations. First, we build upon the CVSS, which remains a widely used standard for assessing vulnerability severity. We adopt relevant CVSS metrics (e.g., attack vector and complexity) while refining others to better suit exploit-level analysis. For example, rather than using CVSS's impact triad (confidentiality, integrity, and availability), which may not directly reflect exploit intent or behavior, we introduce sub-features such as code execution and privilege escalation. These reflect concrete attacker outcomes and align more closely with red team workflows.

Second, we incorporate insights from experienced penetration testers in an industry red team with whom we actively collaborate. Drawing on their operational experience, we identified limitations in existing scoring systems and adjusted our feature definitions accordingly. For instance, the CVSS exploit maturity metric lacks specific guidance for distinguishing between PoC and fully functional exploits. Based on expert input, we introduce more granular sub-features that account for exploit completeness, required setup, and adaptability. We also add metrics such as exploit popularity which captures community adoption and reuse frequency, an aspect practitioners often consider but is missing from CVSS.

Together, these features are designed to reflect not only the theoretical severity of an exploit but also its real-world actionability, supporting both vulnerability triage and exploit selection tasks.


\underline{Attack Vector.}
This feature evaluates whether a vulnerability can be exploited remotely. Remote exploitation significantly broadens the attack surface by allowing attackers to target systems without physical or local network access, making it a crucial factor to consider in exploit assessment.

\begin{table*}[htbp]
\centering
\vspace{-0.1in}
\caption{Feature Values and Criteria}
\small
\label{tab:vul_feature_aggr}
\begin{tabular}{c|c|c}

\toprule
Feature        & Value & Criteria  \\ \midrule
\multirow{2}{*}{Attack Vector}  & Remote  & IsRemote = True\\
                                & Not Remote & Otherwise\\ \midrule
\multirow{2}{*}{Attack Complexity}  &  Low   &  Complexity score $>$ Complexity Threshold \\
                                &  High  &  Otherwise \\ \midrule
\multirow{5}{*}{Impact} & Code Exec. &  Code Exec. = True \\
                        & Privilege Escalation &  Code Exec. = False $\land$ Privilege Escalation = True \\
                        & Info. Leak & Code Exec., Privilege Escalation = False $\land$ Info. Leak = True \\
                        & Bypass & Code Exec., Privilege Escalation, Info. Leak = False $\land$ Bypass = True \\
                        & DoS & DoS = True \\ \midrule
\multirow{3}{*}{Exploit Maturity} & None    & Relevance = False \\
                                  & PoC     & Relevance = True $\land$ ((Availability = False $\land$ Flexibility = False) $\lor$ Functionality = False) \\
                                  & Exploit & Relevance = True $\land$ (Availability = True $\lor$ Flexibility = True) $\land$ Functionality = True \\ \midrule
\multirow{2}{*}{Popularity} &  Low  & Popularity Score $>$ Popularity Threshold \\
                            &  High & Otherwise \\
\bottomrule
\end{tabular}
\vspace{-0.5em}
\end{table*}

\underline{Attack Complexity.}
Attack complexity is assessed through six sub-features. Information dependency considers whether prior knowledge (e.g., credentials) is needed. The attack condition evaluates whether specific configurations are required. Probability dependency examines reliance on uncontrollable factors, such as race conditions. User interaction determines whether user actions (e.g., uploading a file) are necessary. Privilege requirements assess the level of access needed, from none to administrative. Attack evasion checks for techniques to bypass detection. Together, these sub-features provide a detailed view of an exploit's prerequisites and difficulty.

\underline{Impact.}
Impact measures the potential consequences of a successful exploit, ranked by severity. The most critical is code execution, followed by privilege escalation (gaining unauthorized access), information leak (unauthorized data access), bypass (circumventing security mechanisms), and denial of service (disrupting system availability). Feature aggregation prioritizes the most severe potential impact.

\underline{Exploit Maturity.}
Exploit maturity measures the usability and sophistication of exploit code. This is determined through four sub-features. First, relevance evaluates whether the code directly relates to the vulnerability and can at least verify its existence. Second, availability measures whether the exploit code is accessible, such as being publicly hosted in repositories. Third, flexibility examines whether the code allows customization, such as modifying attack targets or goals. Finally, functionality assesses whether the exploit achieves goals beyond verification, such as remote code execution or bypassing authentication. These sub-features collectively provide insight into the practicality and versatility of available exploit code.

\underline{Popularity.}
Popularity is measured using community engagement metrics such as GitHub repository counts, stars, and forks. These metrics collectively indicate how widely a CVE topic is discussed and its relevance within the security community. Stars represent popularity and perceived value, while forks signal active engagement and adaptability. GitHub was selected as the primary data source due to its structured API and rich dataset, providing a reliable and efficient way to assess trends. The popularity score can provide insights of the potential impact and prioritization of a vulnerability, as higher popularity often correlates with broader adoption and community validation.

\begin{tcolorbox}[colback=gray!10, colframe=black, title=Feature Extraction Query (Simplified)]
\small
\begin{center}
    \textbf{Role-play}
\end{center}
You're an excellent cybersecurity expert.
\begin{center}
    \textbf{CoT}
\end{center}
You should analyze the exploit from the following aspects with these steps: \underline{$<$Analysis Steps$>$}

\begin{center}
    \textbf{RAG}
\end{center}
Here are some relevant documentation and code snippets: \underline{$<$Knowledge from RAG$>$}
\begin{center}
    \textbf{Structured Output}
\end{center}
You should always respond in valid JSON format with the following fields: \underline{$<$Format Spec.$>$} 
\\For example, the response should look like this: \underline{$<$Output Example$>$}
\end{tcolorbox}

We utilize LLMs to extract all features except for popularity. To achieve this, we implement an RAG pipeline, which incorporates the collected data as contextual input during feature extraction. Our prompt design strategically employs LLM techniques, such as Chain-of-Thought (CoT) reasoning, to ensure accurate and comprehensive feature extraction. We show some examples of the detailed feature extraction outputs in Appendix~\ref{appendix:output_example}.

For vulnerabilities without publicly available exploits, we assign a low default score. The lack of exploit data reflects the inherent difficulty in exploiting these vulnerabilities, as attackers lack the necessary information or tools to develop effective attacks.



\subsection{Feature Aggregation}
\label{subsec:feature_aggr}
After extracting sub-features, we aggregate them to produce final values for the five key features: Attack Vector, Attack Complexity, Impact, Exploit Maturity, and Popularity. 
The aggregation involves either directly mapping values from sub-features or using criteria and thresholds to determine feature values, as summarized in Table~\ref{tab:vul_feature_aggr}. Finally, we compute an exploit exploitability score to represent the overall exploitability of an exploit and derive vulnerability-level severity by aggregating exploit scores.
\XS{
The aggregation process relies on several sets of weights, which serve as hyperparameters that can be adjusted to suit user needs. We determined the weights used in our experiments from expert elicitation to approximate values recognized by domain experts. We surveyed a panel of experienced pentesters who collaborated with us, asking them to rank the relative importance of features and sub-features. The weights are calculated from these rankings by averaging and normalizing them to approximate weights recognized by domain experts.
}

\underline{Attack Vector.} 
The value of the Attack Vector feature directly corresponds to the IsRemote sub-feature. If the sub-feature indicates that the vulnerability can be exploited remotely (IsRemote = True), the attack vector is labeled as ``True"; otherwise, it is labeled as ``False."

\underline{Attack Complexity.} 
Attack Complexity is derived from a weighted complexity score calculated using six sub-features: information dependency, attack condition, probability dependency, user interaction, privilege required, and attack evasion. Each sub-feature is assigned a weight based on its relative importance to the overall complexity. The weighted score is computed using the formula:
\begin{equation}
    \textit{Complexity Score} = \sum_{i=1}^{6} w_i \cdot f_i
\end{equation}
where $w_i$ represents the weight of sub-feature $i$, and $f_i$ represents the ordinal value of that sub-feature.

If the calculated score exceeds a predefined threshold, the attack complexity is categorized as ``Low"; otherwise, it is categorized as ``High."

\underline{Impact.} 
The Impact feature is determined using a hierarchical evaluation of its four sub-features: code execution, privilege escalation, information leak, and bypass. Each sub-feature is assessed in descending order of severity, and the first applicable sub-feature defines the impact value. For instance, if Code Execution is true, it overwrites the other three sub-features, assigning ``Code Execution" as the impact. In addition, DoS may be added to the Impact if the DoS sub-feature is true, regardless of the other four sub-features.

\underline{Exploit Maturity.} 
This feature is determined by four sub-features: relevance, availability, flexibility, and functionality. Exploits are classified into three levels: None, PoC, and Exploit.None indicates the exploit is irrelevant and cannot verify the vulnerability. PoC applies when the exploit is relevant but fails to perform meaningful attack actions. Exploit is assigned when the exploit is relevant, functional, and either has accessible source code or supports flexible configuration changes (e.g., target, goal), enabling real-world use. This aggregation captures the distinction between trivial code samples and weaponized exploits.

\underline{Popularity.} 
Popularity is derived from metrics such as the number of repositories, stars, and forks associated with the exploit. These values are combined into a Popularity Score using a weighted formula:
\begin{equation}
    \textit{Popularity Score} = w_1 \cdot \textit{Repos} + w_2 \cdot \textit{Stars} + w_3 \cdot \textit{Forks}
\end{equation}
Here, $w_1$, $w_2$, and $w_3$ denote the weights for the respective sub-metrics: number of repositories, stars, and forks. These weights are chosen based on the relative importance of each metric in determining popularity.
If the computed score surpasses a designated threshold, the exploit's popularity is categorized as ``High"; otherwise, it is labeled as ``Low."

\underline{Exploit Actionability Score.}
We calculate the actionability score for the exploit by normalizing the weighted sum of all feature scores.
\begin{equation}
    \textit{Total Score} = \alpha_1 \cdot AV + \alpha_2 \cdot AC + \alpha_3 \cdot I + \alpha_4 \cdot EM + \alpha_5 \cdot P
\end{equation}
\begin{equation}
\label{eq:exp_usability}
    \textit{Act. Score} = \norm{\textit{Total Score}}
\end{equation}
where $AV$, $AC$, $I$, $EM$, and $P$ are the numerical representations of Attack Vector, Attack Complexity, Impact, Exploit Maturity, and Popularity, respectively.
$\alpha_1, \alpha_2, \alpha_3, \alpha_4, \alpha_5$ are weights assigned to each feature.
This score provides a comprehensive measure of an exploit's potential impact and feasibility. Intuitively, the exploitability of exploits of a certain vulnerability can also reflect the severity of the vulnerability.
To calculate the vulnerability severity, we aggregate the exploit scores for all associated exploits using the maximum value: 
\begin{equation} 
\label{eq:vul_severity}
    \textit{Vulnerability Severity} = \max_{j} (\textit{Act. Score}_j) 
\end{equation} where $j$ indexes all exploits related to a specific vulnerability.
The hierarchical feature aggregation process transforms exploit-level sub-features into actionable vulnerability-level metrics, providing a solid foundation for automated and explainable vulnerability and exploit assessment.
At the same time, this process can mitigate potential errors from LLM hallucinations, By considering all associated exploits, inaccuracies in individual exploit features are averaged out, ensuring robust and reliable assessments.

\XS{

\begin{tcolorbox}[colback=gray!10, colframe=black, title=\SysName Output (Simplified)]
    \small
    \begin{center}
        \textbf{Severity Score}
    \end{center}
    The severity score of the vulnerability is: \underline{$<$Severity Score$>$}
    \begin{center}
        \textbf{Exploitability Scores}
    \end{center}
    The exploitability scores of the exploits are:
    \begin{itemize}
        \item Exploit 1: \underline{$<$Exploit 1 Score$>$}
        \item Exploit 2: \underline{$<$Exploit 2 Score$>$}
        \item \ldots
    \end{itemize}
    \begin{center}
        \textbf{Exploit Features}
    \end{center}
    Exploit 1:
    \begin{itemize}
        \item Attack Vector: \underline{$<$Attack Vector$>$}
        \begin{itemize}
            \item \underline{$<$Justification$>$}
        \end{itemize}
        \item Attack Complexity: \underline{$<$Attack Complexity$>$}
        \begin{itemize}
            \item \underline{$<$Justification$>$}
        \end{itemize}
        \item \ldots
    \end{itemize}
    Exploit 2:
    \ldots
\end{tcolorbox}

As shown above, \SysName eventually provides a two-level output for each vulnerability: (1) a vulnerability-level severity score, computed from the highest-ranked associated exploit; and (2) a set of exploit-level exploitability scores and feature values, including attack vector, complexity, impact, maturity, and popularity, each accompanied by textual justifications generated via an LLM. These explanations summarize evidence extracted from exploit code and documentation. The intent is to help users understand not only \emph{what} scores are assigned, but \emph{why} those scores arise, offering transparency for both decision support and learning. 
}


\section{Evaluation}
\label{sec:evaluation}
We evaluate our system to answer the following research questions:
\begin{enumerate}[label=RQ\arabic*., left=0pt, itemsep=0em]
    \item \underline{Exploit Actionability (Section~\ref{subsec:eval_rq1}).} Can \SysName accurately identify actionable exploits? This question focuses on the exploit-level output of the system. We assess whether \SysName can identify effective exploits effectively by comparing its exploit recommendations against ground truth obtained through manual analysis.
    \item \underline{Vulnerability Severity (Section~\ref{subsec:eval_rq2}).} Can \SysName assign severity scores that support effective vulnerability prioritization? We compare \SysName's vulnerability severity scores with those of EPSS using statistical agreement analysis and expert validation to determine whether \SysName offers improved support for practical decision-making.
    \item \underline{Ablation Study (Section~\ref{subsec:eval_rq3}).} How do different LLM backbones and prompting strategies affect feature extraction quality? We examine whether alternative LLMs and prompt designs preserve feature extraction accuracy and consistency.
\end{enumerate}

\subsection{Evaluation Setup}
\subsubsection{Dataset}
\label{subsubsec:dataset}
To evaluate \SysName, we constructed a large and diverse dataset of vulnerabilities derived from real-world penetration testing scenarios. Specifically, we collaborated with industry red team professionals, who provided a list of 655 frequently encountered applications based on their operational experience. We aggregated all known CVEs associated with these applications, resulting in a pool of more than 5,000 vulnerabilities.

\begin{figure}[!htbp]
\vspace{-0.5em}
\centering
\includegraphics[width=0.8\linewidth]{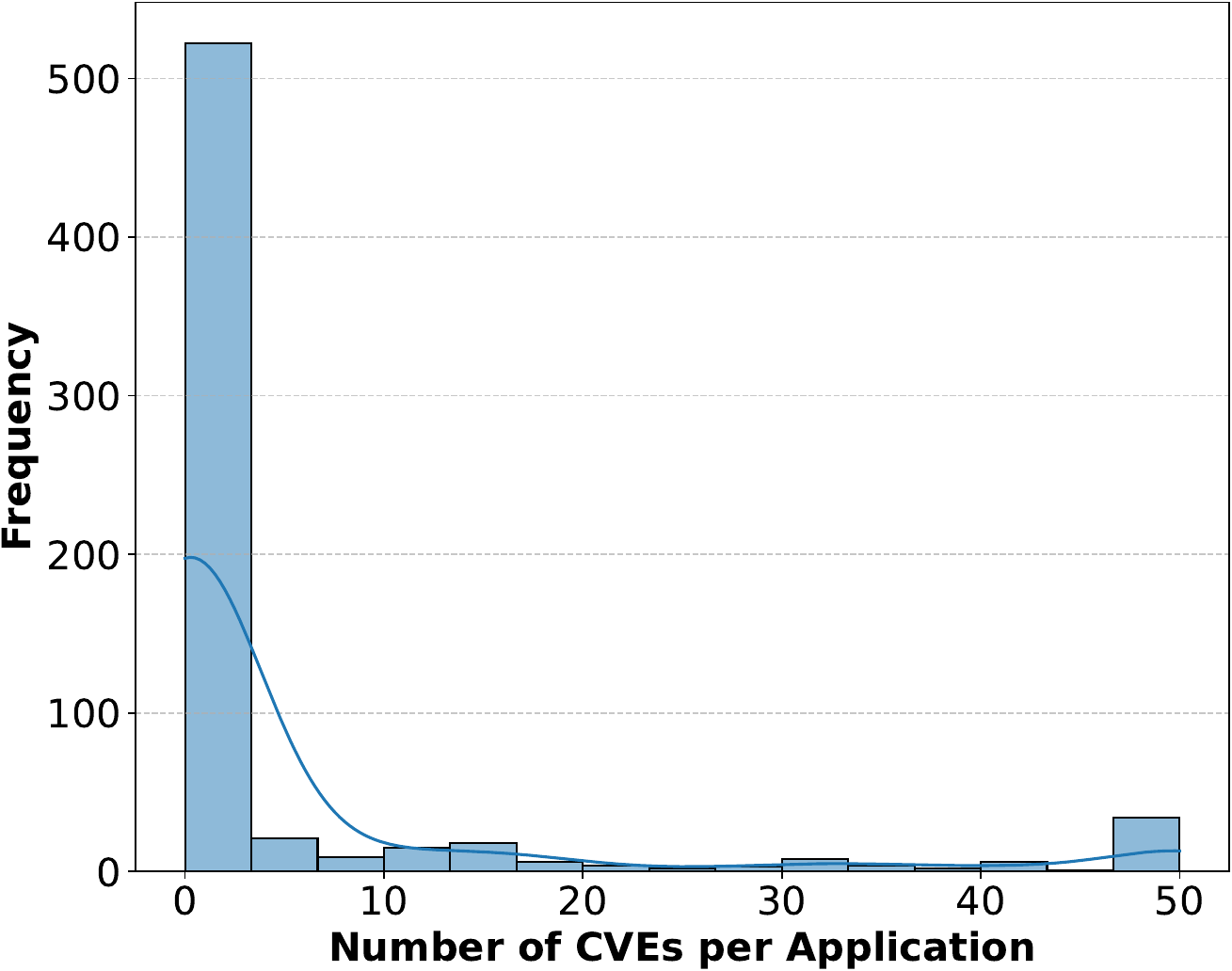}
\caption{Distribution of Number of CVEs per Application}
\label{fig:app_mapping}
\vspace{-0.5em}
\end{figure}

Figure~\ref{fig:app_mapping} shows the distribution of CVEs across these applications. While most applications are associated with only a few vulnerabilities, approximately 15\% (97 out of 655) have more than ten, illustrating the scale and prioritization challenge practitioners face in real environments.

Due to cost and runtime constraints, not all vulnerabilities were used in further in-depth evaluation. Instead, we selected representative subsets of this dataset for different evaluation tasks. These subsets were incrementally sampled to maximize coverage across the following dimensions:



\begin{list}{\labelitemi}{\leftmargin=0.5em}
\item \underline{Vulnerability Types:} We prioritized maximizing coverage across Common Weakness Enumerations (CWEs) to capture a broad range of technical weaknesses and attack scenarios.
\item \underline{Severity Levels:} We ensure balanced coverage across CVSS and EPSS severity ranges to test how well \SysName handles vulnerabilities of varying impact and likelihood.
\item \underline{Mitigation of Prior Knowledge Bias:} To reduce the likelihood that LLMs rely on memorized information from their training data, we focused on vulnerabilities published on or after January 2024. This choice reflects a balance between two goals: mitigating prior knowledge bias, since most models used in our system have training cutoffs in late 2023 (see Table~\ref{tab:llm-models}), and ensuring sufficient availability of public exploit artifacts for analysis. 
While this strategy does not fully eliminate the possibility of memorization, it significantly reduces the influence of prior exposure and allows us to better assess the system's capability to extract features and reason about exploit actionability based on the inputs alone.
\end{list}

This evaluation setup enables a rigorous and realistic assessment of \SysName’s performance, while ensuring the dataset reflects both practitioner needs and research goals.

\subsubsection{Actionability Metrics}
\label{subsubsec:usability_metrics}
To establish a objective standard for evaluating actionability of exploits, we define a set of quantitative metrics. These metrics capture various aspects of the exploit process:

\begin{list}{\labelitemi}{\leftmargin=0.5em}

    \item\underline{Exploit Maturity:} Whether the exploit achieves its intended objective (e.g., code execution, privilege escalation) or it can only serve as a scanner to verify the vulnerability exists. This metric evaluates the practical viability of an exploit.
    
    \item\underline{Completion Time:} The duration required to assess and execute the exploit, starting from its initial inspection and ending with either successful exploitation or a failure conclusion. This metric reflects the efficiency of the exploitation process.

    \item\underline{Number of Errors:} The total number of errors or adjustments needed during the exploitation process. This includes syntax corrections, misconfigurations, and dependencies, providing insight into the exploit’s ease of use.

\end{list}

Together, these metrics provide a comprehensive view of exploit actionability. 

\subsubsection{Environment Setup}
\label{subsubsec:env_setup}
To verify the actionability of exploits, we need to reproduce the vulnerable environments and use the exploits to attempt to exploit these vulnerable environments.
To provide enough isolation for the safety consideration so that our simulated attacks do not cause any unintended damage, we built two virtual machines to serve as the attacker and victim machines.
The simulated vulnerable environments are hosted on a virtual machine with 2 CPU cores and 8 GB RAM, running Ubuntu 22.04 LTS. To avoid interference with the testing process, we have disabled all irrelevant services that allow IO communication, such as SSH.
The attacker machine is also hosted on a virtual machine with 16 CPU cores and 16 GB RAM, running Kali Linux 2024.1.
The victim machine and the attacker machine maintain network connectivity via NAT. This setup ensures the attacker can simulate real-world network conditions when attempting to exploit the vulnerabilities while keeping isolated to the outside Internet.
The assessments are run on an Ubuntu 22.04.3 Linux Server with an Intel(R) Xeon(R) Platinum 8358 CPU @ 2.60GHz and 1.0 TB memory.

We evaluate our framework using a mix of general-purpose and reasoning LLMs. Table~\ref{tab:llm-models} summarizes their key properties.

\begin{table}[htbp] 
\centering 
\caption{Summary of LLM Models}
\label{tab:llm-models}
\begin{tabular}{lcc} 
\toprule 
Model & Context Window & Knowledge Cutoff  \\ 
\midrule 
GPT-4o-mini  & 128k tokens & Oct 2023  \\ 
o3-mini (Reasoning) & 200k tokens & Oct 2023  \\ 
DeepSeek-V3-0324   & 128k tokens & Mar 2025  \\
DeepSeek-R1 (Reasoning)  & 128k tokens & Feb 2024  \\
Kimi-v1  & 128k tokens & Unknown  \\
Kimi-k1.5 (Reasoning) & 128k tokens & Unknown  \\
Llama-3.3-70B-Instruct & 128k tokens & Dec 2023  \\ 
\bottomrule 
\end{tabular} 
\vspace{-0.5em}
\end{table}

\subsection{RQ1. Exploit Actionability}
\label{subsec:eval_rq1}

To evaluate the actionability of \SysName, we conducted a detailed manual verification of exploit artifacts to verify the quality of individual exploits. In this context, exploit actionability refers to the ability of a candidate exploit to achieve its intended goal, such as remote code execution or privilege escalation, with minimal configuration or manual intervention. By manually validating each exploit's functionality, runtime behavior, and execution conditions, we can measure how well \SysName identifies and prioritizes actionable exploits in practice.

\begin{figure}[htbp]
    \centering
    \includegraphics[width=1.0\linewidth]{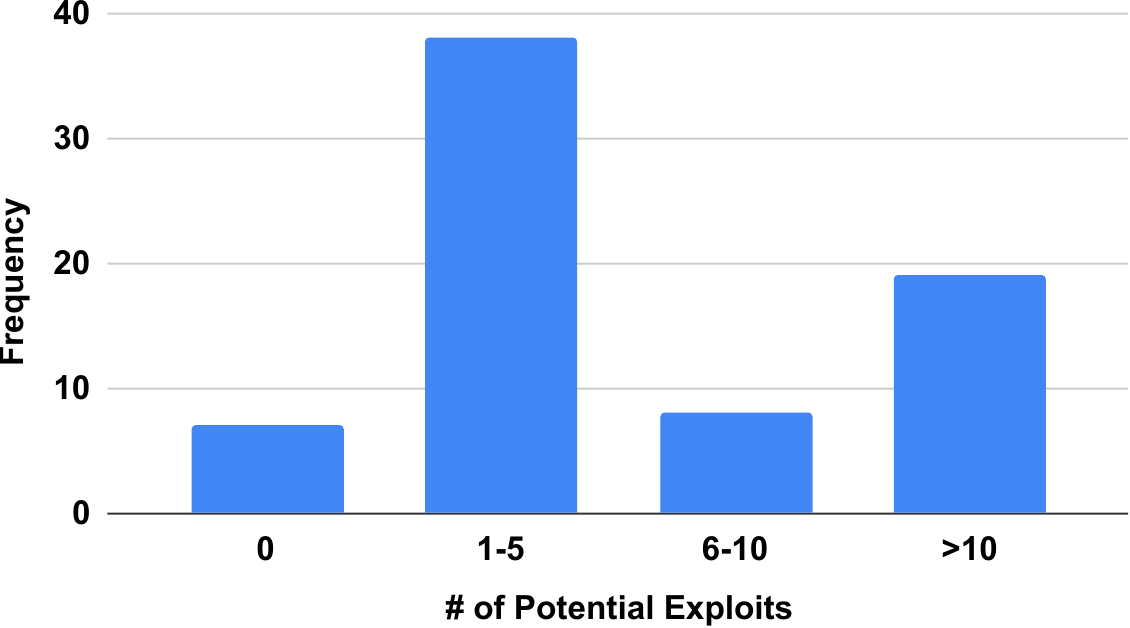}
    \caption{Distribution of Number of Exploits per CVE}
    \label{fig:num_of_repo_histo}
\vspace{-0.1in}
\end{figure}

Since the exploit-level manual analysis is time-consuming, we focused on a testing set of 71 vulnerabilities. 
The testing set was randomly and incrementally sampled from the dataset, ensuring that it included a diverse range of vulnerabilities with varying severities.
Fig.~\ref{fig:num_of_repo_histo} presents a histogram showing the distribution of the number of repositories associated with each vulnerability. Among these vulnerabilities, 7 had no available exploits, 37 had between one and five, 8 had between five and ten, and 19 had more than ten. This distribution underscores the need for effective prioritization mechanisms, as practitioners must often navigate a wide range of exploit options with varying degrees of quality.

\begin{figure}[!htbp]
    \centering
    \includegraphics[width=0.8\linewidth]{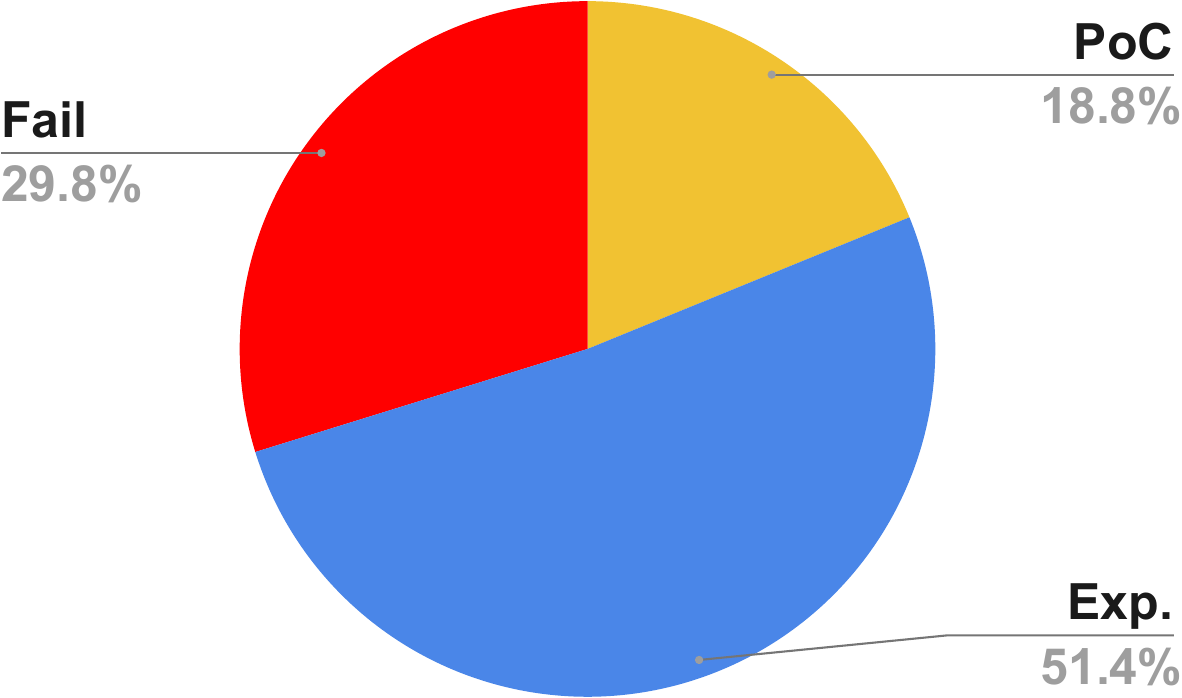}
    \caption{Overall Distribution of Exploit Maturity}
    \label{fig:exp_maturity_dist}
    \vspace{-0.5em}
\end{figure}

\begin{figure}[!htbp]
\vspace{-0.5em}
    \centering
    \includegraphics[width=1.0\linewidth]{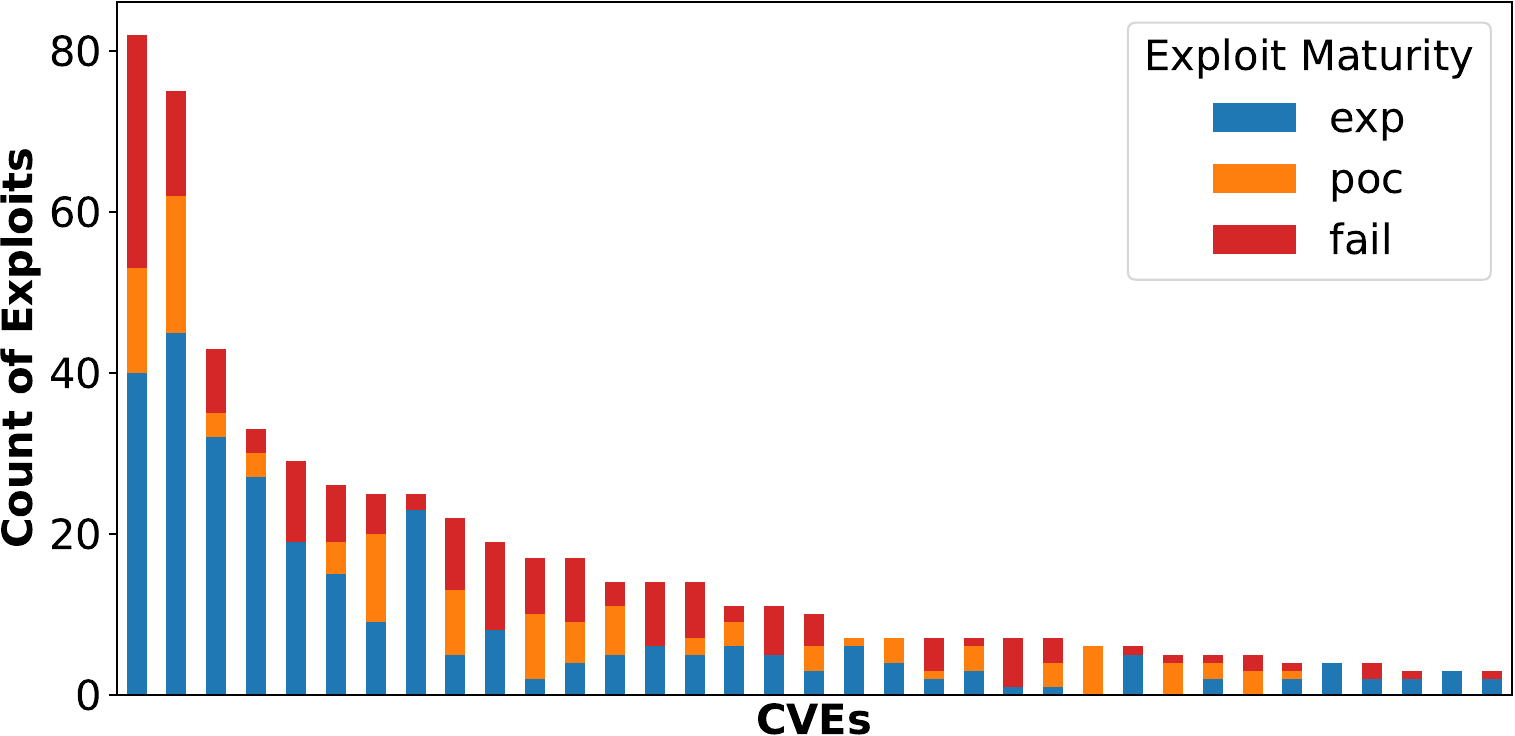}
    \caption{Distribution of Exploit Maturity per CVE}
    \label{fig:cve_level}
    \vspace{-0.5em}
\end{figure}

For the 64 vulnerabilities with available exploits, we manually evaluated a total of 611 potential exploits in the experimental environments described in Section~\ref{subsubsec:env_setup}, using the actionability metrics defined in Section~\ref{subsubsec:usability_metrics}.
The evaluation process involved a consistent evaluator who attempted each exploit, recording to what degree it succeeded in achieving the intended functionality and the time required to achieve success. The evaluator also noted the number of errors encountered during the exploitation process. This manual evaluation provided a comprehensive understanding of the actionability of each candidate, allowing for a detailed analysis of the results.
The evaluation revealed that 36 (5.9\%) of the exploits lacked source code or executable files, providing only descriptive documentation. 115 (18.8\%) exploits were usable as PoCs to verify the existence of vulnerabilities but could not achieve the desired objectives, such as remote code execution. Only 314 (51.4\%) of the evaluated exploits successfully achieved their intended functionality. The remaining 182 (29.8\%) exploits were either non-functional or failed to achieve the intended functionality, indicating a significant number of low-quality or incomplete exploits in the dataset.
Fig.~\ref{fig:exp_maturity_dist} shows an overall distribution of exploit maturity, and Fig.~\ref{fig:cve_level} presents a detailed view for CVEs with three or more exploits. This further underscores the substantial workload required to assess and prioritize the available exploits.
At the vulnerability level, 47 (66.2\%) of the 71 analyzed vulnerabilities had at least one functional exploit, while 54 (76.1\%) had at least a PoC-level exploit available. However, 17 (23.9\%) of the vulnerabilities lacked any functional exploit or PoC, further highlighting the variability in exploit availability and quality.
On average, the successful exploits took 8.5 minutes to complete, while the PoC-level exploits took 9.0 minutes. The number of errors encountered during the exploitation process averaged 1.1 for successful exploits and 2.1 for PoC-level exploits.



We use these manual evaluation results as the ground truth to assess the performance of \SysName's actionability ranking. We use the 47 vulnerabilities that have at least one functional exploit as the dataset. For each vulnerability, \SysName generated an actionability ranking based on scores computed using Equation~\ref{eq:exp_usability}. We compared these rankings with manually verified rankings to assess \SysName’s performance.
Three key metrics were used to evaluate the accuracy and robustness of \SysName in this evaluation: 
\begin{list}{\labelitemi}{\leftmargin=0.5em}
    \item \underline{Top-$k$ Success Rate}, which measures whether at least one exploit achieving the intended functionality appears in the top-$k$ recommendations; 
    \item \underline{Precision@$k$}, which evaluates whether the highest-quality exploit (as determined by manual verification) is included in the top-$k$ recommendations;
    \item \underline{Recall@$k$ for Top-$j$}, which assesses whether at least one of the top-$j$ manually verified exploits is included in the top-$k$ recommendations. 
\end{list}


\begin{table}[ht]
\caption{Exploit Actionability Evaluation Results}
\label{tab:exploit_ranking_metrics}
\centering
\begin{tabular}{lcc}
\toprule
\textbf{Metric}         & \textbf{\SysName}  &  \textbf{Random Select} \\ \midrule
Top-1 Success Rate   & 80.9\%     &       51.4\%        \\ 
Top-3 Success Rate   & 100\%      &       -         \\ 
Precision @ 3 Rate            & 66.0\%      &       -     \\ 
Recall @ 3 for Top-3 Rate            & 78.7\%     &     -       \\ 
\bottomrule

\end{tabular}
\vspace{-0.5em}

\end{table}



Table~\ref{tab:exploit_ranking_metrics} summarizes the results of these evaluations.
The results demonstrate that \SysName effectively prioritizes high-quality exploits. The Top-1 Success Rate shows that 80.9\% of the time, \SysName ranks a successful exploit achieving the intended functionality as its top recommendation. Furthermore, the Top-3 Success Rate indicates that for all evaluated vulnerabilities, at least one functional exploit is included within the top three recommendations, achieving a perfect success rate of 100\%. The results also show that \SysName outperforms random selection, which only achieves a Top-1 Success Rate of 51.4\%.
Besides, Precision@$3$ reveals that in 66.0\% of cases, the highest-quality exploit identified through manual evaluation is among \SysName’s top three recommendations. Finally, the Recall@$3$ for Top-3 metric shows that 78.7\% of the top three manually verified exploits are included within the top three recommendations of \SysName.

These findings validate \SysName’s accuracy and effectiveness in exploit actionability assessment. \SysName reliably identifies and prioritizes high-quality exploits among exploits with various qualities. This not only reduces the time and effort required for manual evaluation but also ensures that practitioners can focus on the most actionable and impactful exploits for their analyses.

\subsection{RQ2. Vulnerability Severity}
\label{subsec:eval_rq2}

We now evaluate \SysName at the vulnerability level to assess how well it supports prioritization decisions when aggregated across multiple exploit candidates. For this evaluation, we selected a representative subset of 292 vulnerabilities and over 1,000 associated exploits using the approach mentioned in Section~\ref{subsubsec:dataset}. While \SysName primarily performs analysis on the exploit level, many security workflows require vulnerability-level severity scores to drive triage, remediation, or risk-based decision-making. To assess the effectiveness of \SysName in this context, we compare its vulnerability severity scores against those generated by EPSS.

EPSS is the most suitable existing baseline for this comparison because it is designed to predict real-world exploitation likelihood based on data-driven modeling. Unlike CVSS, which reflects only theoretical severity and lacks empirical grounding, EPSS incorporates external evidence such as exploitation trends and threat intelligence feeds. Although EPSS does not analyze exploit artifacts directly, it provides the strongest available benchmark for evaluating prioritization effectiveness in operational settings.

We conduct an agreement analysis to quantify how well \SysName's vulnerability scores align with those from EPSS. Agreement between the two can indicate that \SysName captures similar operational signals through artifact-level analysis, while disagreement may suggest that \SysName offers new insights grounded in practical exploitability. This analysis allows us to examine both consistency and divergence, helping us understand whether \SysName produces reasonable vulnerability-level assessments that can serve as a practical complement to existing scoring systems.


\begin{figure*}[!htbp]
     \centering
     \begin{subfigure}[b]{0.48\textwidth}
        \centering
        \includegraphics[width=\linewidth]{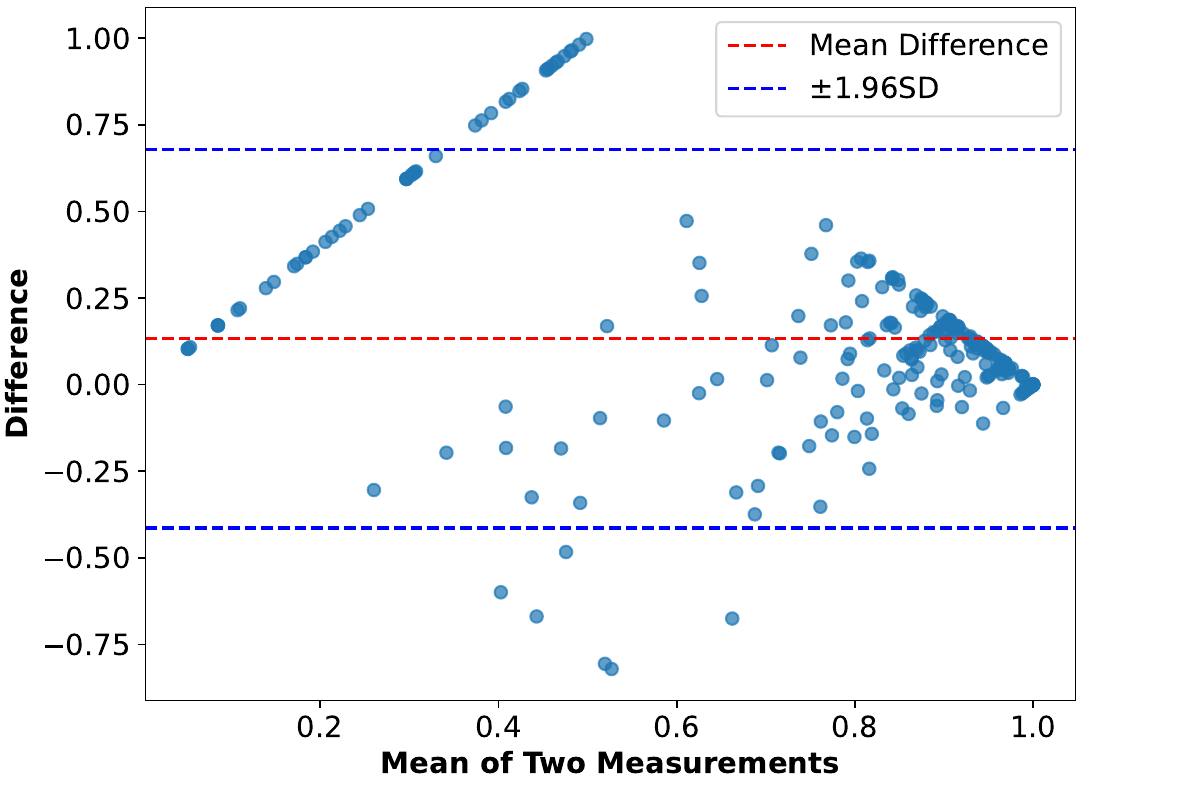}
        \caption{Bland-Altman Plot for All Vulnerabilities}
        \label{subfig:Bland-Altman_all}
     \end{subfigure}
     \hfill
     \begin{subfigure}[b]{0.48\textwidth}
        \centering
        \includegraphics[width=\linewidth]{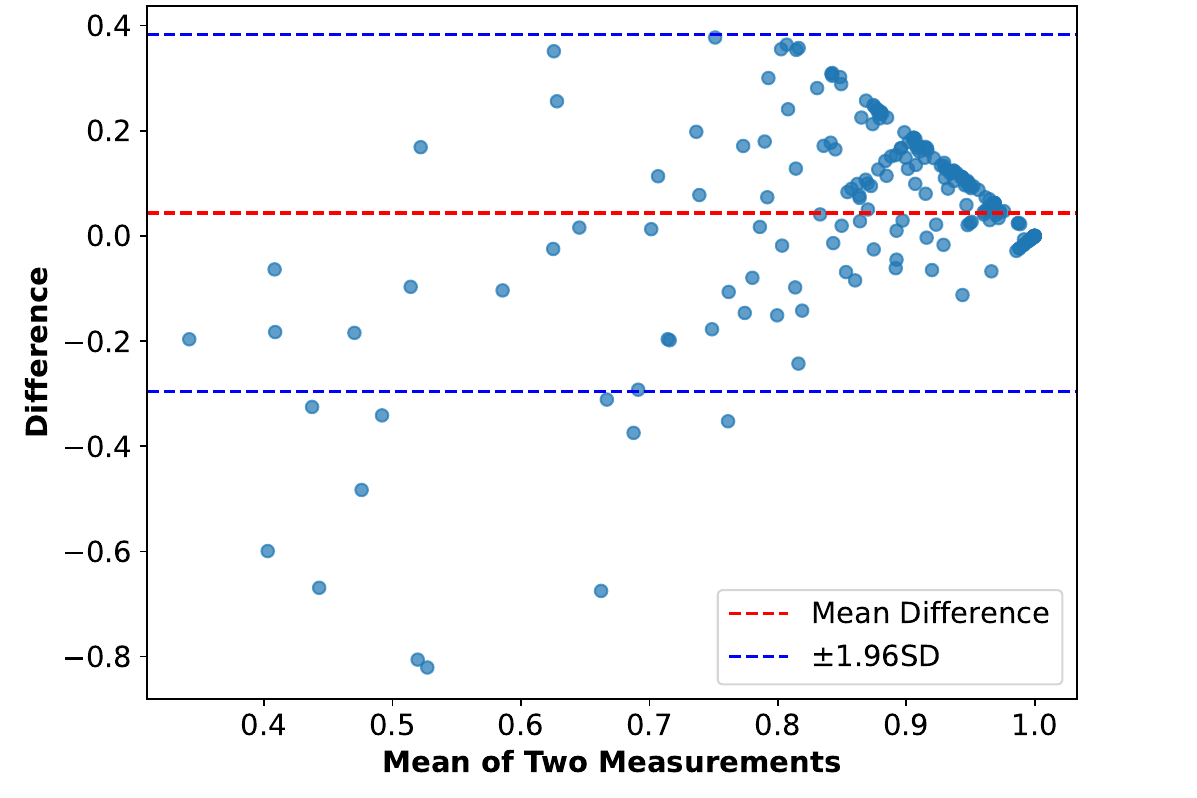}
        \caption{Bland-Altman Plot for Vulnerability with Available Exploits}
        \label{subfig:Bland-Altman_has_code}
     \end{subfigure}
        \caption{Bland-Altman Plots for Vulnerability Severity Comparison}
        \label{fig:Bland-Altman}
        \vspace{-1em}
\end{figure*}

\noindent\underline{Agreement Analysis.}
We evaluate the agreement between \SysName's vulnerability severity scores and EPSS using Bland-Altman analysis, a statistical method that quantifies how closely two sets of measurements align and highlights systematic differences or outliers. Fig.~\ref{fig:Bland-Altman} shows two Bland-Altman plots comparing \SysName's results with EPSS.
Each point in the plot represents a vulnerability, where the x-axis denotes the average of two scores (i.e., \SysName and EPSS), and the y-axis denotes the difference between them. Agreement is considered strong when most differences are small and evenly distributed across the score range. For the full dataset, the mean difference between EPSS and \SysName is 0.13. A total of 32 vulnerabilities (10.96\%) fall outside the 95\% limits of agreement (defined as ±1.96 standard deviations from the mean), meaning that 89.04\% of vulnerabilities show reasonable agreement.

We observe a clear trend in the lower-severity range as shown at the left side of Fig.~\ref{subfig:Bland-Altman_all}. For some vulnerabilities, \SysName consistently assigns lower severity scores than EPSS, especially when no usable exploit is available. In such cases, \SysName assigns a score of 0.0, while EPSS may still assign moderate scores. This reflects a key difference in design: \SysName penalizes vulnerabilities without actionable exploit artifacts, while EPSS relies on statistical signals that may not account for the absence of usable code.

To better isolate the impact of exploit availability, we repeat the analysis on the subset of vulnerabilities that have at least one associated exploit. After removing the 53 vulnerabilities (19.15\% of the dataset) without any exploit data, agreement improves substantially. As shown in Fig.~\ref{subfig:Bland-Altman_has_code}, the mean difference narrows to 0.04, and only 13 vulnerabilities (5.44\%) fall outside the 95\% limits of agreement.

We also observe a second pattern in the high-severity range towards the right side of the plots. When both EPSS and \SysName assign relatively high scores, \SysName tends to produce slightly lower values. This suggests that \SysName differentiates more sharply among high-risk vulnerabilities, assigning lower scores to those that are difficult to exploit despite having high theoretical severity. In contrast, EPSS tends to assign elevated scores more uniformly across this range, possibly due to correlations with metadata features rather than exploit-specific evidence.

These observations highlight a fundamental distinction between the two systems. EPSS provides a statistical estimate of exploitation likelihood using external signals and historical data, while \SysName grounds its assessment in direct analysis of exploit artifacts. As a result, \SysName is less likely to overestimate the risk of vulnerabilities that are severe in theory but impractical to exploit due to missing, incomplete, or unusable exploit code.

\noindent\underline{Outlier Analysis.}
To better understand the discrepancies between \SysName and EPSS, we conducted a detailed case study analysis of outliers identified in the Bland-Altman analysis. Specifically, we analyzed 36 cases with score differences exceeding one standard deviation among vulnerabilities with associated exploits. These outliers fall into two primary categories: vulnerabilities where \SysName assigned higher severity than EPSS and those where it assigned lower severity.

\begin{table}[t]
\caption{Outlier Analysis Summary}
\label{tab:outlier}
\centering
\begin{tabular}{lcc}
\toprule
\textbf{Case}         & \textbf{\# of Case} & \textbf{Accuracy} \\ \midrule
\SysName $\gg$ EPSS      & 9          & 66.7\%                      \\ 
\SysName $\ll$ EPSS      & 27         & 100\%                       \\ 
Overall                & 36               & 91.7\%                      \\
\bottomrule
\end{tabular}

\end{table}

\textit{Case 1}: \SysName determines the vulnerability has a higher severity than EPSS.
In this category, \SysName assigned higher severity to 9 cases (25\%). Among these, three vulnerabilities had working exploits that achieved the attack goals in our experimental environment. EPSS likely assigned lower scores due to factors such as the complexity of exploitation, including kernel-level attacks or evasion techniques. However, \SysName correctly identified automated exploit scripts for these cases, which significantly simplified exploitation, demonstrating its ability to accurately capture actionability in such scenarios. 
Another three cases involved only PoC-level exploits, where the discrepancy arises from EPSS’s underestimation of the severity of these vulnerabilities. Although these exploits were not fully functional, they provided sufficient information to demonstrate the vulnerability's existence and potential impact. \SysName's higher scores in these cases reflect its focus on practical actionability rather than theoretical severity.
The remaining three cases were misclassified by \SysName. One exploit only provided a binary payload, which \SysName could not analyze; another exploit had comprehensive documentation but failed in our manual testing. The last case involved a honeypot exploit that was incorrectly treated as functional by \SysName. These instances represent limitations of our static analysis approach but constitute a small fraction of the dataset.


\textit{Case 2}: \SysName determines the vulnerability has a lower severity than EPSS.
In this category, \SysName assigned lower severity scores than EPSS to 27 cases (75\%). Of these, 12 vulnerabilities did not have any exploit or PoC available through our online search, which naturally led to lower scores under \SysName's exploit-centric assessment. 
The remaining 15 cases shared common characteristics: limited availability of exploit candidates, unclear or missing README files, and minimal code comments, all of which hindered \SysName's ability to analyze the code effectively. 
Additionally, most cases exhibited score differences close to one standard deviation, which aligns with \SysName’s more balanced score distribution compared to EPSS’s heavily skewed distribution. 
Among these 15 cases, six had non-functional exploits upon manual verification, supporting \SysName’s assessment. Another five cases were associated with PoC-level exploits that EPSS scored relatively high, despite their limited functionality. The final four cases involved functional exploits with low actionability due to unclear documentation or manual execution requirements, which \SysName penalized.


Overall, the outlier analysis demonstrates that \SysName provides explainable results for vulnerability severity assessment. This explainability enables \SysName to justify instances where its evaluations diverge from EPSS, supported by exploit-level evidence. These cases demonstrate that \SysName can assess vulnerabilities more accurately than EPSS. By aggregating exploit actionability metrics into vulnerability severity scores, \SysName offers a reliable and interpretable framework that aligns well with established benchmarks while addressing critical gaps in exploit-level analysis.


\noindent\underline{Expert Validation.}
To further evaluate the accuracy and practical relevance of \SysName’s vulnerability severity scores, we conduct an expert validation study with six experienced security professionals from our collaborating institution. Importantly, these experts were not involved in the design, development, or tuning of \SysName. This separation ensures that their feedback is independent and not influenced by prior knowledge of the system’s internal logic or feature choices, thereby reducing potential bias in the evaluation. 


To make the manual validation manageable, we randomly selected 150 vulnerabilities from the larger pool of 5,000+ vulnerabilities analyzed by \SysName. Each expert reviewed a set of 25 vulnerabilities and was shown anonymized, normalized severity scores produced by \SysName, CVSS, and EPSS. For each score, they indicated whether it was \emph{Overrated}, \emph{Underrated}, or a \emph{Match} based on their experience with similar cases.
This expert feedback allows us to directly compare \SysName’s assessments against professional expectations and complements the manual exploit-level validation in Section~\ref{subsec:eval_rq1}.

As shown in Table~\ref{tab:vul_accuracy}, \SysName aligns with expert judgment in 91.3\% of cases, significantly outperforming EPSS (45.3\%) and CVSS (53.3\%). Notably, both EPSS and CVSS exhibit a much higher rate of Overrated assessments (53.3\% and 46.0\%, respectively) compared to only 6.0\% for \SysName. This suggests that existing scoring systems tend to assign higher severity levels than practitioners find appropriate in practice, while \SysName produces scores that are more conservative and consistent with operational expectations.

\begin{table}[ht]
    \caption{Accuracy of Vulnerability Severity Assessment}
    \label{tab:vul_accuracy}
    \centering
    \begin{tabular}{lccc}
    \toprule
    \textbf{Dataset}         & \textbf{\SysName} & \textbf{EPSS} & \textbf{CVSS} \\ 
    \midrule
    Match               &      137 (91.3\%)       &        68 (45.3\%)      &     80 (53.3\%)  \\
    Overrated               &      9 (6\%)       &       80 (53.3\%)      &    69 (46\%)   \\
    Underrated                &     4 (2.7\%)        &      2 (1.3\%)      &     1 (0.7\%)  \\ \bottomrule
    \end{tabular}
    \vspace{-0.5em}
    
\end{table}

Further examination of the Overrated cases from \SysName revealed that the discrepancies were typically not due to errors in feature extraction. Instead, experts indicated that while the technical conditions of the exploit were correctly captured, the overall impact should be reflected in the score. For example, CVE-2023-31419 and CVE-2023-27704 were marked as overrated because their exploitation leads only to denial-of-service (DoS), which experts considered less severe than vulnerabilities enabling remote code execution. These observations underscore the inherent subjectivity in severity assessment, where impact interpretation may vary based on deployment context. Even in such cases, \SysName’s detailed feature-level output still provided valuable and accurate technical signals, offering an objective basis for expert interpretation.

For the few Underrated cases (2.7\%), experts noted that mature exploits were missing from the input set used by \SysName. For instance, CVE-2017-3506 is known to have reliable exploitation pathways, but these were not downloaded by our system during data collection. This limitation reflects the dependency of \SysName on comprehensive exploit discovery, which in rare cases may omit relevant artifacts due to incomplete coverage or metadata inconsistencies.

Overall, this study demonstrates that \SysName produces vulnerability severity assessments that closely align with expert intuition, while maintaining transparency and interpretability through its evidence-driven design.

\begin{table*}[ht]
    \vspace{-1em}
    \caption{Ablation Study on LLM Backbones}
    \label{tab:ablation_llm_backbones}
    \centering
    \begin{tabular}{lccccccc}
    \toprule
    \textbf{Metric}  & \textbf{GPT-4o-mini}  & \textbf{o3-mini} & \textbf{DeepSeek-V3} & \textbf{DeepSeek-R1} & \textbf{Kimi-v1} & \textbf{Kimi-k1.5} & \textbf{Llama-3.3} \\ \midrule
    Avg. Accuracy      & 64.3\%    & 78.7\%     & 76.0\% & 83.3\% & 62.6\% & 67.1\% &     70.5\%       \\ 
    Avg. Variance    & 19.6\%       & 14.7\%     &  17.0\% & 16.7\% & 22.7\% & 21.4\% &     18.6\%       \\  \bottomrule
    \end{tabular}
    \vspace{-1em}
\end{table*}

\begin{table}[ht]
    \caption{Ablation Study on LLM Techniques}
    \label{tab:ablation_llm_techniques}
    \centering
    \begin{tabular}{llcccc}
    \toprule
    \text{Model} & \textbf{Metric} & \textbf{w/ CoT}  &  \textbf{w/o CoT} \\ \midrule
    \multirow{2}{*}{GPT-4o-mini} & Avg. Accuracy      & 64.3\%            &      59.4\%                \\ 
                                 & Avg. Variance    & 19.6\%              &         22.6\%              \\ \midrule
    \multirow{2}{*}{o3-mini} & Avg. Accuracy      & 78.7\%            &      73.3\%                \\ 
                                 & Avg. Variance    & 14.7\%              &         13.5\%              \\
    \bottomrule
    \end{tabular}
    \vspace{-1em}
\end{table}

\subsection{RQ3. Ablation Study}
\label{subsec:eval_rq3}

We conduct an ablation study to evaluate how different components of the LLM-based pipeline in \SysName affect feature extraction performance. Specifically, we examine the influence of (1) LLM backbones and (2) LLM techniques such as chain-of-thought (CoT) reasoning. This analysis assesses the robustness of the pipeline design and its adaptability to different model choices.

Given the high manual effort required to label ground truth at the sub-feature level, we focus on a sample of 10 randomly selected vulnerabilities, which collectively associated with 51 candidate exploits. For each exploit, we manually label six sub-features under the attack complexity feature (defined in Table~\ref{tab:vul_features}), resulting in 306 labeled data points. The feature extraction process is repeated three times for each exploit and each model to measure accuracy and consistency. This setup allows us to perform feature-level evaluation while maintaining manageable annotation overhead.


\noindent\underline{LLM Backbones.}
We compare seven different LLM backbones on their ability to extract structured features from exploit artifacts. 
As shown in Table~\ref{tab:ablation_llm_backbones}, DeepSeek-R1 achieves the highest average accuracy (83.3\%), followed by o3-mini (78.7\%) and DeepSeek-V3 (76.0\%). Even lightweight models such as GPT-4o-mini (64.3\%) and Kimi variants (62.6–67.1\%) demonstrate competitive performance, highlighting the adaptability of our pipeline to different LLM backbones.

It is worth noting that our previous evaluations at the exploit and vulnerability levels were conducted using GPT-4o-mini, selected for its cost efficiency. Despite its relatively modest 64.3\% feature-level accuracy, \SysName achieved strong performance in those evaluations (Section~\ref{subsec:eval_rq1} and~\ref{subsec:eval_rq2}), demonstrating that the system is robust to moderate levels of prediction noise. As long as the LLM outputs are broadly reasonable, our aggregation and ranking procedures yield actionable results. While stronger reasoning models such as DeepSeek or o3-mini can further improve feature-level accuracy, which could potentially benefit expert users who wish to inspect specific feature attributions, lighter models already offer sufficient fidelity for general use cases such as vulnerability triage or automated red teaming.

\noindent\underline{LLM Techniques.}
We also examine how prompting strategies influence feature extraction by comparing performance with and without CoT reasoning. The evaluation is conducted on two OpenAI models: GPT-4o-mini, a lightweighted model without dedicated reasoning support, and o3-mini, a more advanced variant designed to handle reasoning tasks. In both cases, we use task-specific prompts tailored to the exploit assessment. The only difference is that the CoT version includes an explicit step-by-step guide to help the model structure its reasoning. We aim to evaluate whether adding such detailed guidance improves accuracy and stability.

As shown in Table~\ref{tab:ablation_llm_techniques}, CoT improves accuracy for both models. GPT-4o-mini improves from 59.4\% to 64.3\%, and o3-mini improves from 73.3\% to 78.7\%. However, CoT has a differential effect on variance. While it reduces variance for GPT-4o-mini (from 22.6\% to 19.6\%), it slightly increases variance for o3-mini (from 13.5\% to 14.7\%).
This suggests that while CoT helps both models arrive at more accurate predictions, its stabilizing effect is more pronounced for models without built-in reasoning capabilities. In contrast, reasoning-capable models may already follow internal reasoning paths, and externally injected reasoning steps may sometimes introduce misalignment or unnecessary verbosity, leading to higher variability across runs.

Nonetheless, these results demonstrate that CoT is effective even for models designed with reasoning capabilities. This suggests that while such models may be capable of reasoning in general, they are not sufficiently trained for the specific task of exploit assessment. Our task-specific CoT prompts help compensate for this limitation by guiding the model through domain-relevant steps.

\section{Limitation and Future Work}
\label{sec:discussion}

\subsection{Static Analysis Without Execution}
\label{subsec:static_analysis}

\SysName is designed to statically assess the actionability of exploits without executing them. This choice is intentional. Dynamic execution is costly, difficult to scale, and unsuitable for environments that require stealth or reproducibility. Our framework instead aims to approximate the effectiveness of runtime testing through careful extraction and interpretation of structured features. However, static analysis may miss behaviors that only emerge during execution, such as environment-specific interactions or timing-based logic. A promising direction for future work is to integrate selective dynamic validation for ambiguous or high-risk cases, while retaining static analysis as the core strategy for scalability.

\subsection{Dependence on Exploit Availability}
\label{subsec:exploit_availability}

\SysName requires access to exploit artifacts to perform its analysis. Although the system can process both public and private exploit repositories, it cannot evaluate vulnerabilities without available exploit code or accompanying documentation. This means the framework complements, but does not replace, predictive systems that infer exploitability in the absence of known artifacts. Future extensions could explore hybrid designs that combine artifact-driven analysis with predictive modeling to handle vulnerabilities with limited public information.

\subsection{Impact of LLM Evolution}
\label{subsec:llm_evolution}

As large language models continue to evolve, their changing capabilities may influence the performance of \SysName. While stronger models are likely to improve feature extraction accuracy, shifts in model behavior, such as reasoning style or prompt sensitivity, could also introduce unexpected variance or inconsistencies. 
Our evaluation demonstrates that \SysName performs well across different model backbones, suggesting a degree of robustness. However, to ensure long-term reliability, future work could explore adaptive techniques such as prompt tuning and model calibration. Monitoring and benchmarking future LLM versions will also be important to detect performance drift and maintain consistent analysis quality.

\subsection{Scope of Actionability Features}
\label{subsec:scope_actionability}

This work focuses specifically on three dimensions of exploit actionability: availability, functionality, and setup complexity. These dimensions reflect common bottlenecks in exploit deployment and are critical for tasks such as exploit triage and red teaming. However, they do not capture other aspects such as stealth and persistence. These properties are often scenario-dependent and harder to infer from static artifacts alone. Expanding the framework to support broader exploit attributes would require new feature designs, validation criteria, and domain-specific modeling, which we consider promising directions for future exploration.





\section{Conclusion}
\label{sec:conclusion}
In this paper, we presented \SysName, an automated system for actionable exploit assessment. Unlike existing scoring schemes or penetration testing tools that focus on theoretical severity or artifact collection, \SysName prioritizes exploit usability by analyzing whether exploits are available, functional, and require minimal setup. Our system integrates structured feature extraction with task-specific LLM prompting and heuristic scoring to evaluate exploit quality without dynamic execution.

We evaluated \SysName using a large dataset derived from over 5,000 vulnerabilities associated with 655 real-world applications commonly encountered in red teaming operations. Through a combination of manual validation and expert feedback, we demonstrated that \SysName can effectively identify usable exploits and produce vulnerability rankings that align with expert expectations. In particular, the system achieved a 100\% success rate in its top-3 exploit recommendations on a manually verified subset, highlighting its practical value in automated red teaming and vulnerability triage workflows.




\pagebreak

\section*{Ethics Considerations}
Our research does not raise ethical concerns, as it focuses on analyzing day-N vulnerabilities that have already been disclosed to relevant stakeholders. The system aims to help vulnerability managers prioritize remediation using publicly available data, without generating new exploits. All assessments were conducted in isolated environments to minimize real-world risks, ensuring our work has no unintended impact on vulnerable systems.
All evaluators were internal collaborators or colleagues participating voluntarily as part of a system evaluation effort. No personally identifiable or behavioral data was collected, and no compensation was provided. This activity did not constitute formal human-subjects research and did not require IRB approval under our institution’s guidelines.



\bibliographystyle{IEEEtran}
\bibliography{IEEEabrv, ref}

\begin{thebibliography}{10}
\providecommand{\url}[1]{#1}
\csname url@samestyle\endcsname
\providecommand{\newblock}{\relax}
\providecommand{\bibinfo}[2]{#2}
\providecommand{\BIBentrySTDinterwordspacing}{\spaceskip=0pt\relax}
\providecommand{\BIBentryALTinterwordstretchfactor}{4}
\providecommand{\BIBentryALTinterwordspacing}{\spaceskip=\fontdimen2\font plus
\BIBentryALTinterwordstretchfactor\fontdimen3\font minus \fontdimen4\font\relax}
\providecommand{\BIBforeignlanguage}[2]{{%
\expandafter\ifx\csname l@#1\endcsname\relax
\typeout{** WARNING: IEEEtran.bst: No hyphenation pattern has been}%
\typeout{** loaded for the language `#1'. Using the pattern for}%
\typeout{** the default language instead.}%
\else
\language=\csname l@#1\endcsname
\fi
#2}}
\providecommand{\BIBdecl}{\relax}
\BIBdecl

\bibitem{cvss}
\BIBentryALTinterwordspacing
N.~V. Database, ``Common vulnerability scoring system calculator,'' 2024. [Online]. Available: \url{https://nvd.nist.gov/vuln-metrics/cvss/v3-calculator}
\BIBentrySTDinterwordspacing

\bibitem{epss}
\BIBentryALTinterwordspacing
F.~of~Incident~Response and I.~Security~Teams, ``Exploit prediction scoring system (epss),'' 2024. [Online]. Available: \url{https://www.first.org/epss/}
\BIBentrySTDinterwordspacing

\bibitem{wunder2024shedding}
J.~Wunder, A.~Kurtz, C.~Eichenm{\"u}ller, F.~Gassmann, and Z.~Benenson, ``Shedding light on cvss scoring inconsistencies: A user-centric study on evaluating widespread security vulnerabilities,'' in \emph{2024 IEEE Symposium on Security and Privacy (SP)}.\hskip 1em plus 0.5em minus 0.4em\relax IEEE, 2024, pp. 1102--1121.

\bibitem{zhang2023flaw}
S.~Zhang, M.~Cai, M.~Zhang, L.~Zhao, and X.~d.~C. de~Carnavalet, ``The flaw within: Identifying cvss score discrepancies in the nvd,'' in \emph{2023 IEEE International Conference on Cloud Computing Technology and Science (CloudCom)}.\hskip 1em plus 0.5em minus 0.4em\relax IEEE, 2023, pp. 185--192.

\bibitem{allodi2014comparing}
L.~Allodi and F.~Massacci, ``Comparing vulnerability severity and exploits using case-control studies,'' \emph{ACM Transactions on Information and System Security (TISSEC)}, vol.~17, no.~1, pp. 1--20, 2014.

\bibitem{jacobs2023enhancing}
J.~Jacobs, S.~Romanosky, O.~Suciu, B.~Edwards, and A.~Sarabi, ``Enhancing vulnerability prioritization: Data-driven exploit predictions with community-driven insights,'' in \emph{2023 IEEE European Symposium on Security and Privacy Workshops (EuroS\&PW)}.\hskip 1em plus 0.5em minus 0.4em\relax IEEE, 2023, pp. 194--206.

\bibitem{tenable}
\BIBentryALTinterwordspacing
Tenable, ``Tenable vulnerability management,'' 2024. [Online]. Available: \url{https://www.tenable.com/products/vulnerability-management}
\BIBentrySTDinterwordspacing

\bibitem{cvemap}
\BIBentryALTinterwordspacing
P.~Discovery, ``Cvemap,'' 2024. [Online]. Available: \url{https://github.com/projectdiscovery/cvemap}
\BIBentrySTDinterwordspacing

\bibitem{metasploit}
\BIBentryALTinterwordspacing
R.~7, ``Metasploit,'' 2024. [Online]. Available: \url{https://www.metasploit.com/}
\BIBentrySTDinterwordspacing

\bibitem{nuclei}
\BIBentryALTinterwordspacing
P.~Discovery, ``Nuclei,'' 2024. [Online]. Available: \url{https://github.com/projectdiscovery/nuclei}
\BIBentrySTDinterwordspacing

\bibitem{ashiwal2024llm}
V.~Ashiwal, S.~Finster, and A.~Dawoud, ``Llm-based vulnerability sourcing from unstructured data,'' in \emph{2024 IEEE European Symposium on Security and Privacy Workshops (EuroS\&PW)}.\hskip 1em plus 0.5em minus 0.4em\relax IEEE, 2024, pp. 634--641.

\bibitem{milousi2024evaluating}
K.~Milousi, P.~Kiriakidis, N.~Mengidis, G.~Rizos, M.~S. Mazi, A.~Voulgaridis, K.~Votis, and D.~Tzovaras, ``Evaluating cybersecurity risk: A comprehensive comparison of vulnerability scoring methodologies,'' in \emph{Proceedings of the 19th International Conference on Availability, Reliability and Security}, 2024, pp. 1--11.

\bibitem{cvss_exploitability}
\BIBentryALTinterwordspacing
F.~of~Incident~Response and I.~Security~Teams, ``Common vulnerability scoring system v3.0: Specification document,'' 2024. [Online]. Available: \url{https://www.first.org/cvss/specification-document}
\BIBentrySTDinterwordspacing

\bibitem{jacobs2020improving}
J.~Jacobs, S.~Romanosky, I.~Adjerid, and W.~Baker, ``Improving vulnerability remediation through better exploit prediction,'' \emph{Journal of Cybersecurity}, vol.~6, no.~1, p. tyaa015, 2020.

\bibitem{le2022survey}
T.~H. Le, H.~Chen, and M.~A. Babar, ``A survey on data-driven software vulnerability assessment and prioritization,'' \emph{ACM Computing Surveys}, vol.~55, no.~5, pp. 1--39, 2022.

\bibitem{yin2023empowering}
J.~Yin, G.~Chen, W.~Hong, H.~Wang, J.~Cao, and Y.~Miao, ``Empowering vulnerability prioritization: A heterogeneous graph-driven framework for exploitability prediction,'' in \emph{International Conference on Web Information Systems Engineering}.\hskip 1em plus 0.5em minus 0.4em\relax Springer, 2023, pp. 289--299.

\bibitem{fang2020fastembed}
Y.~Fang, Y.~Liu, C.~Huang, and L.~Liu, ``Fastembed: Predicting vulnerability exploitation possibility based on ensemble machine learning algorithm,'' \emph{Plos one}, vol.~15, no.~2, p. e0228439, 2020.

\bibitem{harzevili2023survey}
N.~S. Harzevili, A.~B. Belle, J.~Wang, S.~Wang, Z.~Ming, N.~Nagappan \emph{et~al.}, ``A survey on automated software vulnerability detection using machine learning and deep learning,'' \emph{arXiv preprint arXiv:2306.11673}, 2023.

\bibitem{suciu2022expected}
O.~Suciu, C.~Nelson, Z.~Lyu, T.~Bao, and T.~Dumitraș, ``Expected exploitability: Predicting the development of functional vulnerability exploits,'' in \emph{31st USENIX Security Symposium (USENIX Security 22)}, 2022, pp. 377--394.

\bibitem{metasploit_exploit_ranking}
\BIBentryALTinterwordspacing
R.~7, ``Metasploit documentation - exploit ranking,'' 2024. [Online]. Available: \url{https://adfoster-r7.github.io/metasploit-framework/docs/using-metasploit/intermediate/exploit-ranking.html}
\BIBentrySTDinterwordspacing

\bibitem{fedorchenko2023analytical}
E.~Fedorchenko, E.~Novikova, A.~Fedorchenko, and S.~Verevkin, ``An analytical review of the source code models for exploit analysis,'' \emph{Information}, vol.~14, no.~9, p. 497, 2023.

\bibitem{li2023extracting}
H.~Li, H.~Gao, C.~Wu, and M.~A. Vasarhelyi, ``Extracting financial data from unstructured sources: Leveraging large language models,'' \emph{Journal of Information Systems}, pp. 1--22, 2023.

\bibitem{wiest2024llm}
I.~C. Wiest, F.~Wolf, M.-E. Le{\ss}mann, M.~van Treeck, D.~Ferber, J.~Zhu, H.~Boehme, K.~K. Bressem, H.~Ulrich, M.~P. Ebert \emph{et~al.}, ``Llm-aix: An open source pipeline for information extraction from unstructured medical text based on privacy preserving large language models,'' \emph{medRxiv}, 2024.

\bibitem{wang2025survey}
W.~Wang, Z.~Ma, Z.~Wang, C.~Wu, W.~Chen, X.~Li, and Y.~Yuan, ``A survey of llm-based agents in medicine: How far are we from baymax?'' \emph{arXiv preprint arXiv:2502.11211}, 2025.

\bibitem{lin2025healthgpt}
T.~Lin, W.~Zhang, S.~Li, Y.~Yuan, B.~Yu, H.~Li, W.~He, H.~Jiang, M.~Li, X.~Song \emph{et~al.}, ``Healthgpt: A medical large vision-language model for unifying comprehension and generation via heterogeneous knowledge adaptation,'' \emph{arXiv preprint arXiv:2502.09838}, 2025.

\bibitem{fayyazi2023uses}
R.~Fayyazi and S.~J. Yang, ``On the uses of large language models to interpret ambiguous cyberattack descriptions,'' \emph{arXiv preprint arXiv:2306.14062}, 2023.

\bibitem{purba2023software}
M.~D. Purba, A.~Ghosh, B.~J. Radford, and B.~Chu, ``Software vulnerability detection using large language models,'' in \emph{2023 IEEE 34th International Symposium on Software Reliability Engineering Workshops (ISSREW)}.\hskip 1em plus 0.5em minus 0.4em\relax IEEE, 2023, pp. 112--119.

\bibitem{jensen2024software}
R.~I.~T. Jensen, V.~Tawosi, and S.~Alamir, ``Software vulnerability and functionality assessment using llms,'' in \emph{2024 IEEE/ACM International Workshop on Natural Language-Based Software Engineering (NLBSE)}.\hskip 1em plus 0.5em minus 0.4em\relax IEEE, 2024, pp. 25--28.

\bibitem{zhou2024large}
X.~Zhou, T.~Zhang, and D.~Lo, ``Large language model for vulnerability detection: Emerging results and future directions,'' in \emph{Proceedings of the 2024 ACM/IEEE 44th International Conference on Software Engineering: New Ideas and Emerging Results}, 2024, pp. 47--51.

\bibitem{ahmed2023better}
T.~Ahmed and P.~Devanbu, ``Better patching using llm prompting, via self-consistency,'' in \emph{2023 38th IEEE/ACM International Conference on Automated Software Engineering (ASE)}.\hskip 1em plus 0.5em minus 0.4em\relax IEEE, 2023, pp. 1742--1746.

\bibitem{lewis2020retrieval}
P.~Lewis, E.~Perez, A.~Piktus, F.~Petroni, V.~Karpukhin, N.~Goyal, H.~K{\"u}ttler, M.~Lewis, W.-t. Yih, T.~Rockt{\"a}schel \emph{et~al.}, ``Retrieval-augmented generation for knowledge-intensive nlp tasks,'' \emph{Advances in Neural Information Processing Systems}, vol.~33, pp. 9459--9474, 2020.

\bibitem{wei2022chain}
J.~Wei, X.~Wang, D.~Schuurmans, M.~Bosma, F.~Xia, E.~Chi, Q.~V. Le, D.~Zhou \emph{et~al.}, ``Chain-of-thought prompting elicits reasoning in large language models,'' \emph{Advances in neural information processing systems}, vol.~35, pp. 24\,824--24\,837, 2022.

\bibitem{li2023camel}
G.~Li, H.~Hammoud, H.~Itani, D.~Khizbullin, and B.~Ghanem, ``Camel: Communicative agents for `mind' exploration of large language model society,'' \emph{Advances in Neural Information Processing Systems}, vol.~36, pp. 51\,991--52\,008, 2023.

\bibitem{liu2024we}
M.~X. Liu, F.~Liu, A.~J. Fiannaca, T.~Koo, L.~Dixon, M.~Terry, and C.~J. Cai, ```we need structured output': Towards user-centered constraints on large language model output,'' in \emph{Extended Abstracts of the CHI Conference on Human Factors in Computing Systems}, 2024, pp. 1--9.

\bibitem{collins2004language}
K.~Collins-Thompson and J.~P. Callan, ``A language modeling approach to predicting reading difficulty,'' in \emph{Proceedings of the human language technology conference of the North American chapter of the association for computational linguistics: HLT-NAACL}, 2004, pp. 193--200.

\bibitem{schwarm2005reading}
S.~E. Schwarm and M.~Ostendorf, ``Reading level assessment using support vector machines and statistical language models,'' in \emph{Proceedings of the 43rd annual meeting of the Association for Computational Linguistics (ACL’05)}, 2005, pp. 523--530.

\end{thebibliography}
%



\appendices

\label{sec:appendix}
\section{Pre-processing Details}
\label{appendix:github_filtering}

We utilize publicly available information on Google and GitHub as primary data sources due to their broad coverage of relevant information. However, the search results will often contain a lot of irrelevant data, making it necessary to pre-process the search results to eliminate irrelevant data while retaining meaningful and high-quality information. This ensures that subsequent stages are based on reliable inputs, enhancing the overall efficacy of the system.

\begin{figure*}[htbp]
  \centering
  \includegraphics[width=0.9\textwidth]{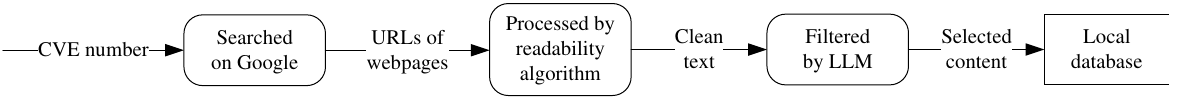}
  \caption{Process of Collecting Data on Google }
  \label{google_process}
\end{figure*}

Google serves as a powerful search engine, providing access to a vast array of web content. However, the variability in search result quality requires a systematic approach to processing and filtering the retrieved data. We begin by extracting webpage content from Google search results and converting HTML documents into text format. Media files, such as images, videos, and audio files, are excluded, as our focus is on analyzing textual data.

To ensure relevance, we apply content-based filtering using the readability algorithm \cite{collins2004language, schwarm2005reading} to remove irrelevant sections of webpages, such as sidebars and advertisements, while keeping critical content like code snippets and documentation as shown in Fig.~\ref{google_process}. Additionally, we utilize an LLM agent with carefully designed prompts incorporating role-playing and few-shot learning techniques. The LLM agent is guided to filter out irrelevant content and emphasize actionable, exploitation-relevant information, such as executable code snippets. We specifically exclude summarized CVE data from vulnerability databases, as such summaries often lack the technical depth necessary for exploitation analysis.

\begin{figure*}[htbp]
  \centering
  \includegraphics[width=0.9\textwidth]{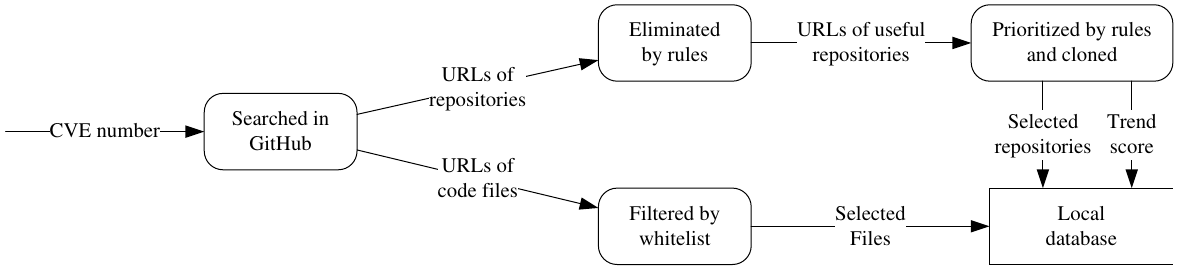}
  \caption{Process of Collecting Data on GitHub }
  \label{github_process}
\end{figure*}

GitHub offers a wealth of repositories containing ready-to-use code, making it a valuable resource for vulnerability assessments. However, the variability in repository quality necessitates a robust filtering and prioritization framework. Our approach incorporates rule-based processing at both the repository and file levels to ensure the relevance and quality of the dataset as outlined in Fig.~\ref{github_process}.

\begin{table}[htbp]
\caption{Heuristic Rule Set for Filtering Repositories}
\label{table_rules}
\centering
\begin{tabular}{cl}
\toprule
Task  & Criteria                                             \\ \midrule
\multirow{3}{*}{Elimination}  & Excessively long descriptions             \\
                       & Abnormal issue counts                     \\
                       & Large number of irrelevant tags \\ \midrule
\multirow{3}{*}{Prioritization}  & Too large or too small sizes              \\
                       & High numbers of forks                     \\
                       & Low numbers of stars                      \\ \bottomrule
\vspace{-0.5em}
\end{tabular}
\end{table}

At the repository level, we implement a two-step process: (1) \emph{Elimination} to exclude low-quality or irrelevant repositories and (2) \emph{Prioritization} to rank higher-quality repositories when multiple candidates are available. Repository quality is assessed using heuristic rules based on metadata attributes, as we outlined in Table~\ref{table_rules}. 
For Elimination, a confidence score is calculated for each repository:
\begin{equation}
    \textit{Confidence Score} = w_1 \times d + w_2 \times i + w_3 \times t
\end{equation}
where $d$ represents the description length, $i$ accounts for issue counts, and $t$ reflects topics associated with the repository.
Repositories scoring below a predefined threshold are excluded.
For Prioritization, repositories are ranked using the quality score:
\begin{equation}
    \textit{Quality Score} =  \frac{ \textit{Stars} \cdot \lambda \cdot \Delta \textit{T}} {\textit{Forks}}
\end{equation}
where $\Delta T$ represents the age of the repository, and $\lambda$ is a normalization parameter to account for differences in creation time. Repositories with higher stars and fewer forks are preferred, as heavily forked repositories often indicate experimental or incomplete scripts unsuitable for direct use.

At the file level, we focus on keeping files that provide meaningful information for vulnerability assessment. These include source code written in high-level programming languages and README files containing detailed descriptions and usage instructions. Conversely, we remove binary files, media files, and configuration files (e.g., XML) that typically lack the content needed for effective analysis. This targeted approach ensures that the kept files are both relevant and suitable for downstream processing.


By meticulously applying these rule-based and LLM-driven filtering and scoring mechanisms, we ensure that the data is not only relevant but also optimized for subsequent analysis. 

\section{\SysName Output Example}
\label{appendix:output_example}

The following examples show outputs of sub-feature extractions. The results are presented in a structured format, including the file reference, the extracted information, and the conclusion drawn from the analysis with confidence levels.

\begin{tcolorbox}[colback=gray!10, colframe=black, title=Information Dependency Analysis]
  \small
  \begin{center}
      \textbf{File Analysis}
  \end{center}
  \begin{enumerate}
    \item CVE-2023-42793.py:47 - Hardcoded credentials (Technique ID 1)
   - Uses hardcoded password "@H454NSec" for created admin accounts
    \item CVE-2023-42793.py:32 - Dynamic credential acquisition (Technique ID 2)
   - Generates random usernames but uses fixed password 
  \end{enumerate}
  \begin{center}
      \textbf{Conclusion}
  \end{center}
True
  \begin{center}
      \textbf{Confidence}
  \end{center}
5
\end{tcolorbox}

\begin{tcolorbox}[colback=gray!10, colframe=black, title=Uncontrollable Preconditions Analysis]
  \small
  \begin{center}
      \textbf{File Analysis}
  \end{center}
  \begin{enumerate}
    \item CVE-2023-42793.py:75 - Hardcoded dependencies (Technique ID 1)
   - Requires specific TeamCity API endpoints (/app/rest/users/id:1/tokens/RPC2)
    \item CVE-2023-42793.py:1 - Third-party service API requirements (Technique ID 2)
   - Requires vulnerable TeamCity REST API to be accessible

  \end{enumerate}
  \begin{center}
      \textbf{Conclusion}
  \end{center}
True
  \begin{center}
      \textbf{Confidence}
  \end{center}
5
\end{tcolorbox}

\begin{tcolorbox}[colback=gray!10, colframe=black, title=Probabilistic Analysis]
  \small
  \begin{center}
      \textbf{File Analysis}
  \end{center}
  \begin{enumerate}
    \item CVE-2023-42793.py:15 - Randomness-dependent execution paths (Technique ID 3)
    - Uses random username generation (1000-9999 range) but this doesn't affect exploit success

  \end{enumerate}
  \begin{center}
      \textbf{Conclusion}
  \end{center}
False
  \begin{center}
      \textbf{Confidence}
  \end{center}
4
\end{tcolorbox}




\section{Additional Evaluation Details}
\label{appendix:evaluation_details}



To complement the Bland-Altman analysis, we calculated additional metrics to quantify the alignment between \SysName and EPSS scores. Table~\ref{tab:vul_agreement} summarizes these metrics, including Pearson Correlation, Spearman Correlation, Mean Absolute Error (MAE), and Root Mean Squared Error (RMSE). For the full dataset, the metrics indicate moderate to strong correlations (Pearson: 0.5772, Spearman: 0.6293) and low error rates (MAE: 0.1979, RMSE: 0.3087). For vulnerabilities with associated exploits, the metrics show improved precision, with lower MAE (0.1208) and RMSE (0.1788), despite a reduced correlation values.
Intuitively, vulnerabilities without available exploits are generally harder to exploit, resulting in lower default scores in \SysName that align with EPSS’s generally lower scores, thereby inflating correlation values.

In summary, \SysName demonstrates a reasonable agreement with EPSS scores, suggesting that it can effectively assess vulnerability severity for most cases. We also performed an outlier analysis to identify vulnerabilities with significant discrepancies between \SysName and EPSS scores, which is detailed in Section~\ref{subsec:eval_rq2} of the main text.

\begin{table}[htbp]
\caption{Vulnerability Severity Agreement Analysis}
\label{tab:vul_agreement}
\vspace{-0.1in}
\centering
\begin{tabular}{lcc}
\toprule
\textbf{Metric}         & \textbf{Value (All)} & \textbf{Value (w/ Exploits)} \\ \midrule
Pearson Correlation      & 0.5772        & 0.4439                      \\ 
Spearman Correlation     & 0.6293        & 0.4924                       \\ 
MAE                  & 0.1979            & 0.1208                    \\ 
RMSE                  & 0.3087            & 0.1788                   \\ \bottomrule
\vspace{-1em}
\end{tabular}
\end{table}

\end{document}